\documentclass[jcph]{apjrnl}
\usepackage{amsmath,psfig,epsfig}

\newcommand{\dv}{\Delta v}
\newcommand{\dr}{\Delta r}
\newcommand{\dt}{\Delta t}
\newcommand{\dx}{\Delta x}
\newcommand{\ip}{{i + \frac{1}{2}}}
\newcommand{\im}{{i - \frac{1}{2}}}
\newcommand{\np}{{n + \frac{1}{2}}}
\newcommand{\gr}{{[i\pm\frac{1}{2}]}}
\newcommand{\gt}{{[n]}}
\newcommand{\usdt}{\frac{1}{\Delta t}}
\newcommand{\ft}{{\tilde{F}}}

\begin{document}

\title{A radiation-hydrodynamics scheme valid from the transport to the diffusion limit.}

\author{E. Audit\ref{sap}, P. Charrier\ref{mab}, J.-P. Chi\`eze\ref{sap} and B. Dubroca\ref{cesta}\\
\additem[sap]{CEA/DSM/DAPNIA Service d'Astrophysique, CEA/Saclay 91191 Gif-sur-Yvette Cedex, France  }\\
\additem[mab]{Applied Mathematics, UMR-CNRS 5466, LRC-CEA M03, 
\* Universit\'e  Bordeaux I, 33405 Talence}\\
\additem[cesta]{CEA-CESTA, BP2, 33114 Le Barp  }
\\E-mail: Edouard.Audit@cea.fr, Pierre.Charrier@math.u-bordeaux.fr, jpchieze@cea.fr, Bruno.Dubroca@cea.fr}

\date{May 7, 2002} 

\maketitle

\begin{abstract}  
  We present  in this  paper the numerical  treatment of  the coupling
  between hydrodynamics and radiative  transfer.  The fluid is modeled
  by classical  conservation laws (mass, momentum and  energy) and the
  radiation  by the grey  moment $M_1$  system \cite{m1}.   The scheme
  introduced  is able to  compute accurate  numerical solution  over a
  broad class of  regimes from the transport to  the diffusive limits.
  We propose an asymptotic  preserving modification of the HLLE scheme
  in order to treat  correctly the diffusion limit.  Several numerical
  results are presented,  which show that this approach  is robust and
  have the  correct behavior in both the  diffusive and free-streaming
  limits.  In  the last numerical example  we test this  approach on a
  complex physical  case by  considering the collapse  of a  gas cloud
  leading to  a proto-stellar  structure which, among  other features,
  exhibits very steep opacity gradients.
\end{abstract}
\begin{keywords}
  Radiative transfer; radiation-hydrodynamics; diffusion limit;
  transport; hyperbolic system; Godunov-type methods
\end{keywords}

\section{Introduction}

Radiation hydrodynamics plays an important role in astrophysics, laser
fusion and plasma physics. For many  years, efforts have been underway
to develop mathematical  models   and  numerical schemes to     obtain
accurate predictions at reasonable computing cost in this domain. \\

One  of   the  main  difficulty   is  to  obtain   accurate  numerical
computations in  the various  regimes that can  be encountered  due to
values  of material  opacities which  can vary  from several  order of
magnitude. This can  be achieved by using the  full radiative transfer
equation  but it  is still  out of  range for  complex  simulations in
particular  in two  or  three spatial  dimensions.   To overcome  this
difficulty several models  have been derived. For large  values of the
material opacity, an asymptotic analysis leads to hyperbolic/parabolic
systems of  equation refered to as the  equilibrium-diffusion limit or
non  equilibrium  diffusion approximation  (\cite{pom},\cite{mihalas},
\cite{lmh}, \cite{dw1}). On the other  hand, for smaller values of the
opacity  the streaming  limit can  be handled  by using  an hyperbolic
system of equations coupling some  closure of the moment system of the
transfer   equation   with   the  hydrodynamic   system   (\cite{pom},
\cite{mihalas},  \cite{lmh},\cite{bal})  .  Unfortunately for  several
real applications,  such as radiative  shocks or star  formation, very
different regimes  are encountered  in the same  simulation.  Coupling
different models  on various zones  introduces large drawbacks  due to
the domain partition and some loss of accuracy in the transitional
zone.\\

From a numerical point  of view high-resolution schemes  are necessary
to  take    into  account  the different    space   and  time  scales.
Godunov-type schemes based on exact or approximate Riemann solvers are
efficient   for  shock     problems  (\cite{godlewski},   \cite{toro},
\cite{roe}).   In   radiation  hydrodynamics, without    source terms,
hydrodynamic and radiation equations decouple so that a Riemann solver
for the entire system can  easily be derived  from Riemann solvers for
the     hydrodynamic       subsystem     and    for     the  radiation
subsystem. Nevertheless, the difficulty  of this approach is to obtain
the correct diffusion limit  for  the scheme  and correct behavior  in
transitional regime.   Let us notice  that the regime at the numerical
level depends  on the ratio of the  mean free path  of photons and the
size of  the local cell.   An other difficulty is  related to the time
scales in  the considered problems. In most  cases radiation has to be
treated implicitly since the time step given by usual CFL condition is
much  too small.  When   possible   the hydrodynamics may  be  treated
explicitly (\cite{dw2}), but  in some problems  the time step given by
the  CFL  condition  on  the   hydrodynamic  subsystem  is still   too
restrictive and hence a fully implicit approach is necessary. \\

In this  paper we propose new  model and numerical  scheme to describe
radiation hydrodynamics in a wide range of regimes.  The equations for
the  fluid are  the classical  conservation laws  (mass,  momentum and
energy). The radiation  is described by the grey  moment system of the
transfer equation (radiative energy and radiative flux) with the $M_1$
closure introduced in (\cite{m1}).  This closure, based on the minimum
entropy  principle,  is  able  to  describe large  anisotropy  of  the
radiation  while  giving  also   the  correct  diffusion  limit.   The
Eddington factor  is a  non constant function  of the reduce  flux and
therefore the  radiation subsystem  is a nonlinear  hyperbolic system.
Here, these  equations are written in the  comoving frame, introducing
some non  conservative coupling terms. A Riemann  solver is developped
by  performing  a wave  decomposition  neglecting the  nonconservative
coupling terms  and the source terms,  which is trivial  since the two
subsystems  then decouple.  Afterwards  the nonconservative  terms are
reintroduced in  the scheme, by  using in the radiation  equations the
velocity given  by the hydrodynamic solver. This  approach, natural in
the streaming regime, can be  improved to capture the diffusion limit.
For  that  purpose the  Riemann  solver  for  the $M_1$  subsystem  is
modified  to take  into account  the stiff  relaxation term  along the
ideas  developped  in   (\cite{sjin}).   With  this  modification  our
approach is  able to correctly treat  the streaming regime  as well as
the  diffusion regime  with  the same  solver.   Finally as  radiative
shocks  or star formation  simulations need  high resolution  and very
different time scales we  describe the implementation of our radiation
hydrodynamics solver in a moving grid-time implicit framework.

The outline  of this paper is  as follows.  In the  second section the
equations of radiation hydrodynamics with the $M_1$-closure are given.
In  section 3  we  present  the numerical  methods  for solving  these
equations.   The  actual implementation  is  detailed  in  section 4.  
Numerical  experiments are  shown  in section  5,  first on  radiative
shocks to  demonstrate the validity  of our method in  various regime,
then on the collapse of a gas cloud leading to a stellar-like
structure. The last section is the conclusion of the paper.\\

\section{The physical model}

The equations of grey  radiative hydrodynamics under the assumption of
local thermodynamic  equilibrium can be found  in (\cite{mihalas}). We
have choosen in  this work to use the radiative  equation in the fluid
(or comoving) frame \nolinebreak : 

\begin{alignat}{6}
\partial_t \rho  & \ + \ && \nabla . [\rho u] && =  && 0  \label{hydro1}\\
\partial_t  \rho u    & \ + \ && \nabla . [\rho u\otimes u + P\mathbf{I}] &&=  
&& && \frac{\kappa}{c} \rho F_r + F \label{hydro2}\\
\partial_t  E    & \ + \ && \nabla . [u(E + P)] && =  -&& \kappa\rho c(a_r T^4 - E_r) 
+(  && \frac{\kappa}{c} \rho F_r + F).u \label{hydro3}\\
\partial_t E_r & + \nabla . [E_r u] \ + \ && \nabla . F_r +  
\mathbf{P}_r : \nabla u 
&& = && \kappa\rho c(a_r T^4 - E_r) \label{rayo1} \\
\partial_t F_r & + \nabla . [F_r u]   \ + \ && c^2 \nabla .\mathbf{P}_r + (F_r . \nabla) u && 
= - && \kappa\rho c F_r \label{rayo2} 
\end{alignat}

\noindent where  $\rho$ is the matter  density, u the  velocity, E the
total matter energy density, P  and T the pressure and the temperature
of  the material,  F the  force  density, $E_r$  the radiative  energy
density,  $F_r$  the  radiative  flux,  $\mathbf{P}_r$  the  radiative
pressure tensor and $\kappa$ the mean grey opacity.

The  first part of  the system  \eqref{hydro1}-\eqref{hydro3} describe
the    evolution    of    the    matter    and    the    second    one
\eqref{rayo1}-\eqref{rayo2} the evolution of the radiation.  These two
fluids  are coupled  by the  exchange of  energy and  momentum through
respectively the $\kappa\rho c(a_r T^4 - E_r)$ and $\kappa \rho F_r/c$
terms.

In order to close system \eqref{hydro1}-\eqref{rayo2} it is necessary
to give two equations  of state for  the  flow and for  the radiation.
For  the flow fields we assume  a $\gamma$-law $p  =  (\gamma -1) \rho
\varepsilon$   (with  $\gamma=5/3$  in   our  applications).   For the
radiation   fields we need   a closure relation   giving the radiative
pressure $\mathbf{P}_r$ in  term of the  radiative energy $E_r$ and of
the radiative flux  $F_r$.  A rather simple closure  is to assume that
the   radiative  flux  is     isotrope which  leads  to    the closure
$\mathbf{P}_r =  \frac{1}{3}    E_r$.  Such an    approximation  is by
construction  not  very  well suited  to   model flow containing large
anisotropy  in the  radiative  flux which is   generally the case  for
radiative shocks.  We will therefore use the closure given by the $M1$
model \cite{m1}.  In this model, the radiative pressure is given by:

\begin{equation}
\mathbf{P}_r= \mathbf{D} E_r
\end{equation}
where the Eddington tensor   {\bf D} is defined by:

\begin{align}
\mathbf{D}& = \frac{1-\chi}{2} \mathbf{I} + \frac{3\chi - 1}{2} n\otimes n \\
\chi      & = \frac{3 + 4|f|^2}{5+2\sqrt{4-3|f|^2}} 
\end{align} 

{\bf  I}  is the identity matrix,  $\chi$  the Eddington factor,
$f=\frac{F_r}{cE_r}$  the reduce flux and  $n =  f / |f|$  is
a unit vector aligned with the radiative flux.\\

Let  us remark    that  we can    write the   conservation   of  total
(i.e. radiative and material) momentum and energy. Neglecting terms in
$(u/c)^2$ and terms involving the fluid acceleration this leads to the
following equations

\begin{eqnarray}
\partial_t (\rho u + \frac{1}{c^2} F_r)+\nabla .[\rho u\otimes u + P\mathbf{I}
+ \mathbf{P}_r + u\frac{F_r}{c^2}] &=& F - (\frac{F_r}{c^2}.\nabla)u  \label{totmom}\\
\partial_t (E + E_r + u\frac{F_r}{c^2}) + \nabla . [u(E + E_r + P + P_r ) + F_r ]  &=&  
  F.u   \label{toten}
\end{eqnarray}


Some authors  make the choice  to discretize directly the  total mass,
momentum and energy conservation equations with either the material or
radiative momentum and energy equations  to close the system.  In this
paper we prefere to  work with the system (\ref{hydro1})-(\ref{rayo2})
where  the flow and  the radiation  play a  more symmetric  role.  The
coupling between the flow and  the radiation is done by the relaxation
source  terms and  by the terms  introduced by  the
comoving frame. The mathematical analysis of the hyperbolicity of this
system do not take into  account the source terms. The Jacobian ${\cal
A}$ of the system (\ref{hydro1})-(\ref{rayo2}) 
can be written 
\begin{equation}
  \label{jacful}
  {\cal A} = 
  \begin{pmatrix}
    {\cal A}_{hh} && {\cal A}_{hr} \\ 
    {\cal A}_{rh} && {\cal A}_{rr} 
  \end{pmatrix}
\end{equation}
where the blocks are associated with the partitioning of unknowns
between hydrodynamic variables (denoted $h$) and radiative variables
(denoted $r$).
The hydrodynamic diagonal block ${\cal A}_{hh}$ is classical and writes
\begin{equation}
  \label{jachh}
  {\cal A}_{hh} = 
  \begin{pmatrix}
    0 & 1 & 0 \\
    (\gamma - 3) \frac{u^2}{2} & -(\gamma - 3) u & \gamma - 1 \\
    -u \left ( H - (\gamma - 1) \frac{u^2}{2} \right ) &
    H - (\gamma - 1) u^2 & u \gamma
  \end{pmatrix}
\end{equation}
where $H = \epsilon + p / \rho + u^2 / 2$ is the total enthalpy of the
flow. Since the radiative variables do not appear in the
left-hand-side of the hydrodynamic equations we have 
\begin{equation}
  \label{jachr}
  {\cal A}_{hr} = 
  \begin{pmatrix} 
    0 & 0 \\
    0 & 0 \\
    0 & 0
  \end{pmatrix},
\end{equation}
 
The radiative diagonal block ${\cal A}_{rr}$ for the M1 subsystem in
the comoving frame depends on the velocity of the flow and writes
\begin{equation}
  \label{jacrr}
  {\cal A}_{rr} = 
  \begin{pmatrix} 
    u & 1 \\
    c^2 (\chi(f) - \chi'(f) f)& u + c \chi'(f)
  \end{pmatrix}.
\end{equation}
Finally, the off-diagonal block due to the influence of hydrodynamic
on the left-hand side of the radiative equations can be stated
\begin{equation}
  \label{jacrh}
  {\cal A}_{rh} = \frac{1}{\rho}
  \begin{pmatrix} 
    -u (E_r + P_r) & (E_r + P_r) & 0 \\
    -u F_r & F_r & 0
  \end{pmatrix}.
\end{equation}
 Since the jacobian matrix ${\cal A}$ is a block lower triangular
matrix  its spectrum can 
easily  be  computed  and  is  composed  of  the  eigenvalues  of  the
hydrodynamic subsystem, i.e. $u-a$, $u$, $u+a$, where $a$ is the speed
of   sound   and   of   $\lambda_{-}+u$  and   $\lambda_{+}+u$   where
$\lambda_{-}$   and   $\lambda_{+}$  are   the   eigenvalues  of   the
M1-radiation subsystem in the laboratory frame (cf.  \cite{m1} and
section  3.3). We remark 
that the nonconservative coupling terms ($\mathbf{P}_r : \nabla u$ and
$F_r . \nabla$ u) appear only in the off-diagonal block ${\cal
A}_{rh}$ and thus do not  influence the signal velocities of the whole
system.

\section{Numerical treatment}

The  nonconservative   form  of  the   radiation-hydrodynamics  system
prevents using  classical Godunov type  scheme without care.   In this
paper  we focus  on  some  applications such  as  radiative shocks  or
collapse of  a gas cloud where  various regime are  encountered in the
same simulations.

Our  numerical  strategy is  the  following.  The conservative  system
obtained by dropping the  nonconservative coupling terms is dicretized
by  a  Godunov-type   scheme  based  on  a  Riemann   solver  and  the
nonconservative terms are discretized using the same values as in this
Riemann solver {\em at the same time step}. As we will see in the next
sections the  construction of the Riemann solver  for the conservative
system  is quite  simplified  by the  lower  triangular structure  and
merely reduces  to the construction  of two decoupled  Riemann solvers
for the hydrodynamic and radiation subsystems.
     
The relaxation terms introduce an other kind of difficulty.  For large
value of the  opacity they are stiff and  an asymptotic analysis shows
that  the total  energy is  governed  by a  diffusion equation.   This
regime  is commonly  referred  to as  the equilibrium-diffusion  limit
(\cite{mihalas},\cite{pom}).   To capture this  regime with  the above
system  we  must  treat   carefully  the  relaxation  terms  using  an
asymptotic  preserving  scheme   following  the  ideas  introduced  in
(\cite{sjin}).   This  will be  detailed  in  section 3.3.2.   Finally
equations (\ref{hydro1}),  (\ref{totmom}) and (\ref{toten})  show that
in  a stationnary  shock  between  two states  where  radiation is  in
equilibrium  with material,  the two  states  satisfy Rankine-Hugoniot
relations involving  both the  radiative and material  quantities.  We
will  show  in  sections  5.1   and  5.2  that  our  numerical  scheme
\*satisfies these  relations even when there is  large energy transfer
between matter and
radiation.\\

In many astrophysical situations   physical quantities vary over  many
orders of  magnitude  in  a  very  steep  way.   To  deal which   such
situations, the adaptive grid   technic  proposed by Dorfy  and  Drury
\cite{dd} (thereafter DD)   is very well  suited.   We  have therefore
choosen to  implement our  radiative-hydrodynamical  scheme on such  a
grid.    However, the numerical  schemes presented  below  can also be
applied  on a static  eulerian grid  which would  be better suited for
multi-dimensionnal simulations.\\

Let us first start with some general remarks and notations. Space is
discretized into computational zones centered on points $r_i$. The
grid interfaces are located at $r_{i\pm 1/2}$.  The volume of a zone
is given by:
\begin{equation*}
\dv_i = 
\begin{cases}
\frac{4}{3}\pi(r_{\ip}^3 - r_{\im}^3)  &\mbox{in spherical geometry}\\
r_{\ip} - r_{\im} = \dr_i              &\mbox{in slab geometry (assuming a unit surface)}
\end{cases}
\end{equation*}

Subscripts ($i, \ip, i+1...$)  always stand for the space coordinates,
while  superscipts ($n,  \np...$)  indicate time.   Variables at  time
$\np$ are defined by:

\begin{equation*}
X^{\np}_{i} = (1-\alpha)X^{n}_{i} + \alpha X^{n+1}_{i} \;\;\;\;\;\; \alpha\in [0,1]
\end{equation*}

\noindent for $\alpha = 0$ the scheme will be explicit and for $\alpha = 1$ 
it will be  fully implicit. We also  introduce the following  notations
which are useful to have more compact equations:
\begin{equation*}
X_\gr = (X_\ip - X_\im) \;\;\;\;\mbox{and}\;\;\;\; X^\gt = (X^{n+1} - X^{n})
\end{equation*}
\noindent
Subscript and superscript are always {\it distributive}:
\begin{equation*}
(XY)_\gr^n = X^n_\ip Y^n_\ip - X^n_\im Y^n_\im
\end{equation*}

\subsection{The grid}
In the  following, we briefly  summarize  the adaptive  grid method of
DD. The grid concentration is defined by:

$$ n_i = \frac{L}{r_\ip-r_\im}$$

\noindent  where $L$  is a  typical length.  If $L$  is a  constant, a
uniform concentration corresponds  to a uniform grid and  if $L=r_i$ a
uniform concentration  corresponds to  a logarithmic grid.  This later
case is useful for spherical geometry.

The basic  idea  of DD  is  to demand that   $n$ is proportional to  a
resolution function  $\mathcal{R}$.  This means  that there will  be a
large number of gridpoints wherever the  resolution function is large.
The resolution  function depends on the  nature of the  problems being
solved, but it is most of the time a  function of the magnitude of the
gradient of the physical quantities in order to have a large number of
gridpoints where there are steep features in the flow.

In  order for  the grid  to be  computationally tractable  it  is also
necessary to ensure that adjacent zones have roughly the same size and
that the  grid adaptation time is  not too small  compared to physical
time scale.   These two conditions are fulfilled  using the following
equation:

\begin{align}
\label{raideur}
\hat{n}_i   & = n_i - \alpha_g(\alpha_g+1)(n_{i-1} - 2n_i + n_{i+1}) \\
\label{taug}
\tilde{n}_i^{n} & = \hat{n}_i^{n} + \frac{\tau_g}{\Delta t} (\hat{n}_i^{n} - \hat{n}_i^{n-1})
\end{align}

\noindent where $ \alpha_g$ is the grid rigidity and $\tau_g$ the grid
time-scale.    Equation   (\ref{raideur})   ensures  that   the   grid
concentration does not vary from more than a factor $ (\alpha_g + 1) /
\alpha_g$ between to adjacent zones and equation (\ref{taug}) set the
time-scale on which the grid can readjust itself.

The final form of the grid equation is then:
\begin{equation}
\label{equrada}
\frac{\tilde{n}_{i+1}}{\mathcal{R}_{i+1}}= \frac{\tilde{n}_i}{\mathcal{R}_i} 
\end{equation}

This equation  depends both on the  grid coordinates $r_i$  and on the
physical quantities entering the resolution function $\mathcal{R}$.

\subsection{The hydrodynamics}\label{section_hydro}

In one  dimension and  taking into account  the moving  grid equations
\eqref{hydro1}-\eqref{hydro3} can be written:
\begin{equation}
\label{hydroS}
\begin{cases}
\begin{array}{lllll}
\partial_t \rho S    &+& \partial_r [\rho S (u-u_g) ] & = & 0\\
\partial_t \rho S u   &+& \partial_r [\rho S u(u-u_g) + PS]  & = & P\partial_r S 
+ S(\frac{\kappa}{c} \rho F_r + F)\\
\partial_t  E S   &+& \partial_r [ES(u-u_g) + PSu] & =  & - S \kappa\rho c(a_r T^4 - E_r) \\
& & & & -  S (\frac{\kappa}{c} \rho F_r + F).u
\end{array}
\end{cases}
\end{equation}
where $u_g$ is the  grid velocity and the surface $S$ equals
$1$ in slab geometry and $4\pi r^2$ in spherical geometry.

The left hand  side of system (\ref{hydroS})  is  an hyperbolic system
for the variables $\rho S$, $\rho S u$ and $S  E$. Now, by introducing
the moment vector ${\cal U} = (\rho, \rho u, E)$ and the physical flux
vector   ${\cal F(U)} =  (\rho u,  p  + \rho  u^2,   u (E  + P))$, the
hyperbolic part  of  system (\ref{hydroS}) can   be written under  the
following form:
\begin{equation}
  \label{defhyperpart}
  \partial_t \left ( S {\cal U} \right ) 
  + \partial_r \left [ S ( {\cal F(U)} - u_g \, S {\cal U} ) \right ] = 0.
\end{equation}

As the physical flux vector is a degree one homogeneous vector valued
function ($S {\cal F(U)} = {\cal F(S U)}$), if we set $\hat{\cal U} =
S {\cal U}$, the hyperbolic part \eqref{defhyperpart} can be rewritten
as:
\begin{equation}
  \label{defhypart1}
  \partial_t \hat{\cal U}  + \partial_r \left ( {\cal F(\hat{U})} 
    - u_g \, \hat{\cal U} \right ) = 0.
\end{equation}

The system \eqref{defhypart1} can be solved using an exact Riemann
solver (Godunov scheme):
\begin{equation}
\label{god1}
\usdt  \Bigl( \hat{\cal U} \Delta r  \Bigr)^{\gt}_i 
+ \Bigl({\cal F(\hat{U}^*)} - u_g \, \hat{\cal U}^* \Bigr)_{\gr}^{n+\frac{1}{2}} 
= 0
\end{equation}

The state $\hat{U}^*_{\ip}$ is the solution on the line $r/t=0$ of the
Riemann problem between the cell $i$ and $i+1$ involving the following
initial conditions:
\begin{equation}
  \begin{cases}
    \hat{U}(0,r) = S_{\ip} U_{\ip}^- & \text{for } r < 0, \\
    \hat{U}(0,r) = S_{\ip} U_{\ip}^+ & \text{for } r > 0,
  \end{cases}
\end{equation}

where  $S_{\ip}$  is  the  surface associated  with  interface  $\ip$.
$U_{\ip}^-$ and $U_{\ip}^+$ are the trace of $U$ from the cell $i$ and
$i+1$. For a first order scheme (no Van-Leer extrapolation) $U_{\ip}^-
= U_i$ and  $U_{\ip}^+ = U_{i+1}$. (for details  about the solution of
the Riemann problem see \cite{toro})

We can note that we have used the solution of Riemann problem based on
${\cal F(\hat{U})}     - u_g  \,   \hat{\cal  U}$  instead   of ${\cal
F(\hat{U})}$   flux. In fact,   by  the  Galilean invariance of  Euler
equations,  the solution of Riemann  problem  for ${\cal F(\hat{U})} -
u_g \, \hat{\cal U}$ flux on $r  / t = 0$ is  the same as the solution
of Riemann problem for ${\cal F(\hat{U})}$ flux on $r / t = u_g$.

Unfortunately, the uniform flow is not a numerical solution of the
scheme \eqref{god1}. To satisfy this property, the mesh geometry must
be solution of a balance equation called geometric conservation law
(GCL):
\begin{equation}
\label{gcl1}
\usdt  \Bigl( S \Delta r \Bigr)^{\gt}_i 
- \Bigl(S u_g \Bigr)_{\gr}^{n+\frac{1}{2}} 
= 0.
\end{equation}

Of course, this relation is false for the spherical case. Moreover, if
the product $S_i \Delta r_i$ has been replaced by the volume of a cell
$\dv_i$,  a new  numerical scheme  is obtained  which satisfies  a GCL
balance law:
\begin{equation}
\label{gcl2}
\usdt  \Bigl( \dv \Bigr)^{\gt}_i 
- \Bigl(S u_g \Bigr)_{\gr}^{n+\frac{1}{2}} 
= 0,
\end{equation}
\begin{equation}
\label{god2}
\usdt  \Bigl( {\cal U} \dv \Bigr)^{\gt}_i 
+ \Bigl({\cal F(\hat{U}^*)} - u_g \, \hat{\cal U}^* \Bigr)_{\gr}^{n+\frac{1}{2}} 
= 0.
\end{equation}

The uniform flows are numerical solutions of the scheme
\eqref{god2}. Hence, the whole numerical scheme about system \eqref{hydroS}
can be written now under the following form:
\begin{equation}
\label{equdens}
\usdt  \Bigl( \rho \dv \Bigr)^{\gt}_i 
+ \bigl(\rho^*(u^* - u_g)\bigr)_{\gr}^{n+\frac{1}{2}} = 0
\end{equation}

\begin{equation}
\label{equmom}
\begin{split} 
\usdt \Bigl(\rho u \dv \Bigr)^{\gt}_{i} &
+ \bigl(\rho^* u^* (u^* - u_g) + P^* \bigr)_{\gr}^{\np}  = \\
 & \Bigl(\partial_r S \; P  \dr + (\frac{\kappa}{c} \rho F_r + F)\dv \Bigr)_{i}^{\np}
\end{split}
\end{equation}

\begin{equation}
\label{equener}
\begin{split} 
\usdt \Bigl( E \dv \Bigr)^{\gt}_{i} &
+ \bigl(E^* (u^* - u_g) + P^* u^* \bigr)_{\gr}^{\np}  = \\
 & \Bigl( \bigl( \kappa\rho c(a_r T^4 - E_r) 
+  (\frac{\kappa}{c} \rho F_r + F)u \bigr) \dv \Bigr)_{i}^{\np}
\end{split}
\end{equation}

\subsection{Solving for the radiative transfer}

To  simplify  the presentation,  we  will  first  study the  radiative
transfer in  a frozen medium  and in a  slab geometry. In  this case,
system \eqref{rayo1}-\eqref{rayo2} gives 

\begin{align}
\partial_t E_r & + \;\partial_x  F_r  = \sigma c (a T_m^4 - E_r)
\label{equ1_1d}\\ 
\partial_t F_r & + c^2\partial_x \mathbf{P}_r = -\sigma c Fr \label{equ2_1d}
\end{align}

\noindent  where we have defined $\sigma = \kappa \rho$. Using the $M_1$ closure, the radiative pressure is given by : 
\begin{equation}
\mathbf{P}_r = \chi E_r  = \frac{3 + 4f^2}{5+2\sqrt{4-3f^2}} E_r  \qquad \mbox{where } \qquad f=\frac{F_r}{cE_r}
\end{equation} 

The previous sytem is a hyperbolic one which can be writen under the
following form:
\begin{equation}
\partial_t {\cal U} + \partial_x \cal{F(U)} = \cal{S(U)}
\end{equation}

with
\begin{equation}
\cal{U}    = \begin{pmatrix} E_r \\ F_r \end{pmatrix} \;\;\;\;\;\;\;\;
\cal{F(U)} = \begin{pmatrix} F_r \\ c^2 \chi E_r \end{pmatrix} \;\;\;\;\;\;\;\;
\cal{S(U)} = \begin{pmatrix} \sigma c (a T_m^4 - E_r) \\ -\sigma c F
\end{pmatrix}  
\end{equation}

${\cal U}$ is the moment vector, ${\cal F}$ the physical flux vector
and ${\cal S}$ the source term vector. The Jacobian matrix $J$ of this
system is a function of $f$ and $c$ and  is given by:
\begin{equation*}
 \mathbf{J}({\cal U})  = \frac{\partial \cal{F(U)}}{\partial \cal{U}} =
  \begin{pmatrix} 
    0 & 1 \\
    c^2 (\chi(f) - \chi'(f) f)& c \chi'(f)
   \end{pmatrix}
\end{equation*}

\noindent The eigenvalues, $\lambda^-,\lambda^+ (\lambda^- < \lambda^+$)
 of the matrix 
\* $\mathbf{J}$ can now be  calculated:
\begin{subequations}       
\begin{align}    
  \lambda^- =  \frac{\chi'(f) - \sqrt{\chi'(f)^2 - 4 f \chi'(f) + 4 \chi(f)}}{2} \ c,
    \\
  \lambda^+ =  \frac{\chi'(f) + \sqrt{\chi'(f)^2 - 4 f \chi'(f) + 4 \chi(f)}}{2} \ c.
\end{align}
\end{subequations}       

The   eigenvalues, $\lambda^-$ and   $\lambda^+$,  of $\mathbf{J}$ are
plotted in figure [\ref{figvp}-A].  At  radiative equilibrium (when $f =
0$), the eigenvalues are equal to $\pm  c/\sqrt{3}$, which is a direct
consequence of the isotropy of underlying distribution function of the
\* photons.
   On the other hand, in  the case  of extreme non-equilibrium,
when $f=1$  (resp.  $f=-1$ ), the  eigenvalues are degenerated and are
equals  to   $c$ (resp.    $-c$).   This    case corresponds  to   the
free-streaming limit   where all  the  photons move  towards the  same
direction. It  is  important to note  that  in the  case of a constant
Eddington factor ($\chi = 1/3$), the eigenvalues are also constant and
equal to  $\pm  c/\sqrt{3}$, which means that  a  light  front can not
propagate at the speed of light, as it should.

\setlength{\unitlength}{1.0cm}
\begin{figure}[httb]
\begin{center}
\begin{picture}(12,8)
\put(0.0,0.0){
\psfig{file=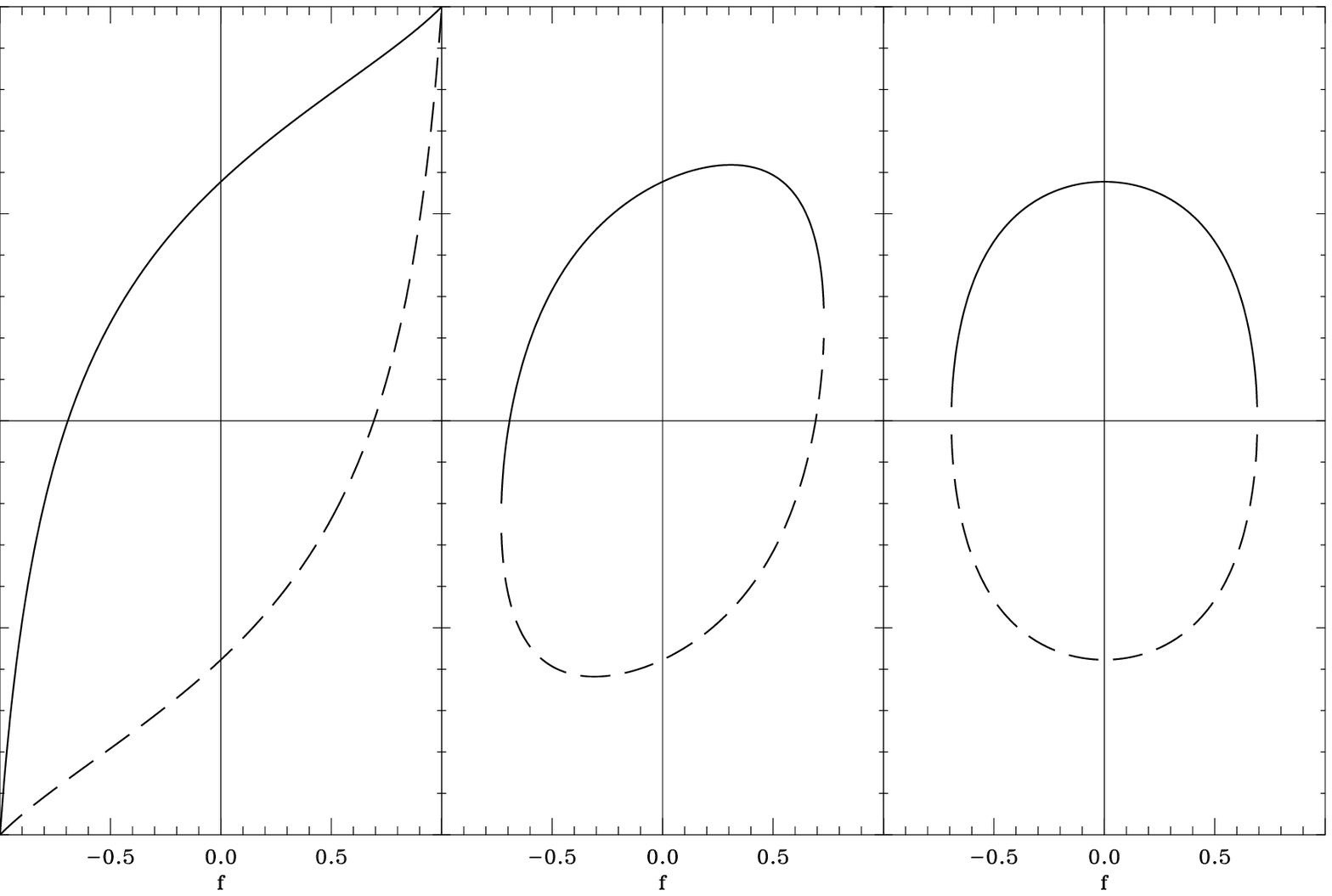,height=7cm,width=14cm}}
\put( 1.0,6.0){A}
\put( 6.0,6.0){B}
\put(10.5,6.0){C}
\end{picture}
\end{center}
\caption{\it eigenvalues  of $M_1$, $\frac{\lambda^+}{c}$  (full line)
  and    $\frac{\lambda^-}{c}$     (dashed    line)    according    to
  $f=\frac{F_r}{cE_r}$,   the   normalized  flux   for   $\epsilon  =   1$
  (A),$\epsilon = 0.5$ (B) and $\epsilon = 0$ (C), }
\label{figvp}
\end{figure}

\subsubsection{Numerical approximation of $P_1$ and $M_1$ by the HLLE  approximate Riemann solver}

The system (\ref{equ1_1d})-(\ref{equ2_1d})  can be discretized using a
Godunov method  based on the  HLLE approximate \*Rieamann  solver (see
\cite{hlle}).  Hence, we need to find an estimation of the largest and
smallest  physical  signal-velocities in  the  exact  solution to  the
Rieamann  problem. This  estimation  involves the  eigenvalues of  the
Roe's  linearization for  the  $M_1$ model.   The  $M_1$ Roe's  matrix
$\mathbf{J_{\ip}}$ between $i$ and $i+1$ states can be written as:

\begin{equation*}    
  \mathbf{J_{\ip}} = 
  \begin{pmatrix} 
    0 & 1 \\
    c^2 (\chi_a - \chi'_m f_a)& c \chi'_m
  \end{pmatrix},
\end{equation*}
\noindent where
\begin{align*}
f_a = \frac{f_i + f_{i+1}}{2}, &&  \chi_a = \frac{\chi(f_i) + \chi(f_{i+1})}{2}, &&
\chi'_m = 
\begin{cases}
  \frac{\chi(f_{i+1}) - \chi(f_{i})}{f_{i+1}-f_{i}} & \text{if $f_{i} \ne f_{i+1}$}, \\
  \chi'(f_{i}) & \text{else}.
\end{cases}
\end{align*}

The eigenvalues, $\lambda^-_{\ip},\lambda^+_{\ip} (\lambda^-_{\ip} < \lambda^+_{\ip} $) of the
matrix $\mathbf{J_{\ip}}$ can now be  calculated:
\begin{subequations}       
\begin{align}    
  \lambda^-_{\ip} =  \frac{\chi'_m - \sqrt{{\chi'_m}^2 - 4 f_a \chi'_m + 4 \chi_a}}{2} \ c,
    \\
  \lambda^+_{\ip} =  \frac{\chi'_m + \sqrt{{\chi'_m}^2 - 4 f_a \chi'_m + 4 \chi_a}}{2} \ c.
\end{align}
\end{subequations}       

Once  knowing the  numerical signal  velocities $\lambda^{\pm}_{\ip}$,
the  numerical flux function  ${\cal F}_{\ip}$  between cells  $i$ and
$i+1$ can be written under the following form:
\begin{align}
\nonumber
{\cal F}_{\ip} = \frac{a^+{\cal F}_{i} - a^-{\cal F}_{i+1} 
                          + a^+a^- ({\cal U}_{i+1}-{\cal U}_{i})} {a^+-a^-}
\;\;\;\; \mbox{where} \;\;\;\; 
\begin{cases}
a^+ = max(\lambda^+_{i+1},\lambda^+_{\ip},0), \\
a^- = min(\lambda^-_i,\lambda^-_{\ip},0).
\end{cases}
\end{align}

\noindent     
The system (\ref{equ1_1d})-(\ref{equ2_1d}) can then be discretized in
conservative form: 

\begin{align}
\frac{E^{n+1}_{r,i} - E^{n}_{r,i}}{\Delta t} & 
      + \frac{F^{\np}_{r,\ip} - F^{\np}_{r,\im}}{\Delta x}
      = \sigma^{\np}_{i} c (a_r T_m^4  - E^{\np}_{r,i})\\ 
\frac{F^{n+1}_{r,i} - F^{n}_{r,i}}{\Delta t} &
      + c^2\frac{P^{\np}_{r,\ip} - P^{\np}_{r,\im}}{\Delta x}
      = -\sigma^{\np}_{i} c F^{\np}_{r,i}
\end{align}

\subsubsection{Scheme modification to enforce asymptotic limit}
\label{modif}

As we mentionned before, one of the important difficulties of
radiative \*transfer is to be able to deal with very different
regimes. It is therefore important to ensure that our scheme has the
correct limiting behavior both in the free-streaming and diffusive
regimes.

In order to illustrate these points in a simple maner we will consider
wave speeds equal to the speed of  light (i.e. $a^+ = -a^- = c$) and a
constant Eddington factor  $\chi = 1/3 $ (resp.  $1$) in the diffusive
(resp. free-streaming) regime. The discretized equations then take the
following form:

\begin{align}
\nonumber
\frac{E^{n+1}_{r,i} - E^{n}_{r,i}}{\Delta t} & 
      + \frac{F^{\np}_{r,i+1} - F^{\np}_{r,i-1}}{2\Delta x}
      - \frac{c}{2\Delta x}(E^{\np}_{r,i+1} -2E^{\np}_{i}+E^{\np}_{r,i-1})
      = \sigma^{\np}_{i} c (a_r T_m^4  - E^{\np}_{r,i})\\ 
\nonumber
\frac{F^{n+1}_{r,i} - F^{n}_{r,i}}{\Delta t} &
      + c^2\chi\frac{E^{\np}_{r,i+1} - E^{\np}_{r,i-1}}{2\Delta x}
      - \frac{c}{2\Delta x}(F^{\np}_{r,i+1} -2F^{\np}_{r,i}+F^{\np}_{r,i-1})
      = -\sigma^{\np}_{i} c F^{\np}_{r,i}
\end{align}

\noindent  the modified equations actually solved by the numerical scheme are therefore:

\begin{align}
\label{equ1b}
\partial_t E_r & + \partial_x  F_r -\frac{c}{2} \Delta x \partial_x^2 E_r 
                               = \sigma c (a T_m^4 - E_r)\\
\label{equ2b}
\partial_t F_r & + c^2\chi\partial_x E_r - \frac{c \Delta x}{2}
\partial_x^2 F_r = -\sigma c F_r 
\end{align}
  
\noindent{\bf Free-streaming regime}

\noindent Let us  first consider the case of  light freely propagating
in vaccum (i.e. $\sigma = 0.$)  with all the photons going in the same
direction (i.e. $F  = c E$). The resulting equation  for the energy of
the radiation is then:

\begin{equation}
\partial_t E  + c \partial_x E_r - \frac{c\Delta x}{2}\partial_x^2 E_r = 0. 
\end{equation}

\noindent This equation is the  exact advection equation with an added
numerical  diffusion term.   The single  mode solutions  of  the exact
equation are of the form: $E_r = E_{r,0} \cos(2\pi/\lambda(ct-x))$. For this
mode  and under  CFL  condition the  variation  of energy  due to  the
diffusion term can be estimated: 

\begin{equation}
\nonumber
\frac{\Delta E_r}{\dt} = \frac{4\pi^2}{2}E_{r,0} \frac{c\dx}{\lambda^2} \Rightarrow 
\frac{\Delta E_r}{E_{r,0}} = \frac{4\pi^2}{2} \frac{c\dt\dx}{\lambda^2} \sim \Bigl(\frac{\dx}{\lambda}\Bigl)^2
\end{equation}

\noindent If  the wave-length of the perturbation  is properly sampled
by the grid the diffusion term will always remain negligible. As could
be  expected, there  will  be  some diffusion  for  very sharp  energy
profiles.  In  actual  simulations,  this diffusion  will  be  further
reduced by taking into account the proper waves speed and we will show
in the next section that it remains acceptable.\\

\noindent{\bf Diffusive regime}

Let us now  examine the diffusive regime.  In  this case, the temporal
variation  of   the  radiative   flux  can  be   neglected.   Equation
(\ref{equ2b}) then gives at first order:

\begin{equation}
F_r \simeq - \frac{c}{3\sigma}\partial_x E_r
\end{equation}

\noindent inserting the previous expression into equation (\ref{equ1b}) one gets:

\begin{equation}
  \partial_t E_r  - \frac{c}{3\sigma}(1+\frac{3 \sigma \Delta x}{2})\partial_x^2 E_r
  = \sigma c (a T_m^4 - E_r)
\end{equation}

The  equation obtained  for  $E_r$  is indeed  a  diffusion equation.  
Furthermore, if the grid is sufficiently fine so that $\sigma \Delta x
<< 1$, the diffusion coefficient is the right one. One the other hand,
if the grid does not sample  the photon mean free path ($\sigma \Delta
x  >> 1$), the  diffusion coefficient  is then  dominated by  a purely
numerical term and  it is therefore much larger  than it should.  With
this simple analysis  it is possible to explain  some of our numerical
results. For a coarse grid, the effective diffusion coefficient is too
large  and the  thermal  precursor is  therefore  longer.  This  wrong
behavior in  the diffusive regime is  not a caracteristic  of the HLLE
scheme we  have used.  In any  Godounov type method  the divergence is
discretized  as the  sum of  a centered  term and  numerical diffusion
term.  This last  term is absolutly necessary to  ensure the stability
of the scheme.   Generally this term is proportionnal  to $\Delta x^n,
(n=1,2)$.   When the mean  free path  of photon  is much  smaller than
$\Delta x$,  the numerical diffusion  term can become dominant,  as we
have just seen.

In order to have a scheme which \* reproduces correctly the diffusion
limit, it is therefore necessary to explicitly take into account the
smallness of the photon mean free path.  We present below such a
scheme.

We first write  the system of equation (\ref{equ1_1d})-(\ref{equ2_1d})
in a dimensionless way.  For that purpose we consider a characteristic
length $l$ and we defined the following quantities:

$$\epsilon = \frac{1}{\sigma l} \qquad \tilde{t} = \frac{c t}{\sigma l^2} \qquad 
\tilde{x} = \frac{x}{l} \qquad \tilde{F_r} = \frac{F_r}{c \, \epsilon}$$

\noindent the  system (\ref{equ1_1d}-\ref{equ2_1d}) can then be written:
\begin{align}
\partial_{\tilde{t}} E_r & + \partial_{\tilde{x}} \tilde{F_r}  
= \frac{(a T_m^4 - E_r)}{\epsilon^2} \label{equ1_mod}\\
\partial_{\tilde{t}} \tilde{F_r} & + \frac{1}{\epsilon^2}\partial_{\tilde{x}} \chi E_r
 = -\frac{1}{\epsilon^2} \tilde{F_r} \label{equ2_mod}
\end{align}

The difficulty raised by the diffusive limit is highlighted in
equation (\ref{equ2_mod}).  When $\epsilon$ is very small, the
hyperbolic system (\ref{equ1_mod})-(\ref{equ2_mod}) becomes very stiff
because of the second term of equation (\ref{equ2_mod}) (the jacobian
condition number $\to \infty$ when $\epsilon \to 0$).  An interesting
way to overcome this difficulty is to transform the previous system in
a non-stiff hyperbolic system with stiff source term \cite{sjin}.
Equation (\ref{equ2_mod}) can be written :

\begin{equation}
  \label{equ2_mod2} 
  \partial_{\tilde{t}} \tilde{F_r}  + \partial_{\tilde{x}} \chi E_r = 
  -\frac{1}{\epsilon^2}
  (\tilde{F}+(1-\epsilon^2)\partial_{\tilde{x}}  \chi E_r)
\end{equation}

The  new  system  to  be  integrated  is  now  composed  of  equations
(\ref{equ1_mod})  and (\ref{equ2_mod2}).  In  order to  integrate this
system the  hyperbolic terms (to the  left of the equal  sign) will be
treated by the Godounov type method presented above, whereas the other
terms  (to the  right of  the equal  sign) will  be treated  as source
terms.  Which means that they will be discretized in a purely centered
manner.       When       $\epsilon      =      1$       the      sytem
(\ref{equ1_mod})-(\ref{equ2_mod2})    is     identical    to    system
(\ref{equ1_1d})-(\ref{equ2_1d})   and  gives  the   correct  transport
regime.   But  when  $\epsilon  \rightarrow  0$, the  source  term  in
equation  (\ref{equ2_mod2})  becomes dominant  and  gives the  correct
radiative flux  in the diffusion  limit.  Furthermore, because  of the
change  of $F$  into $\tilde{F_r}$,  the numerical  diffusion  term in
equation (\ref{equ1_1d})  has been  multiplied by $\epsilon$  and will
therefore always remain small.

The hyperbolic system to be solved  is now:

\begin{equation}
\label{sysm}
\begin{cases}
\partial_{\tilde{t}} E_r  + \partial_{\tilde{x}} \tilde{F_r} = 0\\
\partial_{\tilde{t}} \tilde{F_r}  + \partial_{\tilde{x}} \chi E_r = 0
\end{cases}
\end{equation}

As  previously  done, we  will  integrate  the  above system  using  a
approximate HLLE Riemann solver. The Jacobian of the system (\ref{sysm}) is:

$$
\mathbf{J} = \begin{bmatrix}
               0            &     1 \\
               \chi - f\chi' & \epsilon \chi'
             \end{bmatrix}      
\qquad \mbox{ with } \;\; f=F_r/cE_r, \;\;\chi=\chi(f) \;\;\mbox{ and }\;\; \chi'= 
\frac{d\chi(f)}{df}
$$

\noindent The  eigenvalues, 
$\tilde{\lambda}^+$ and $\tilde{\lambda}^-$ ($\tilde{\lambda}^+ >
\tilde{\lambda}^-$), of the Jacobian are given by:

$$
\label{vp2}
\tilde{\lambda}^\pm =  \frac{\left(\epsilon \chi'\pm \sqrt{(\epsilon \chi')^2 
+4(\chi - f\chi') } \right)}{2}
$$

\begin{figure}
\hspace{-0.0cm}\epsfig{file=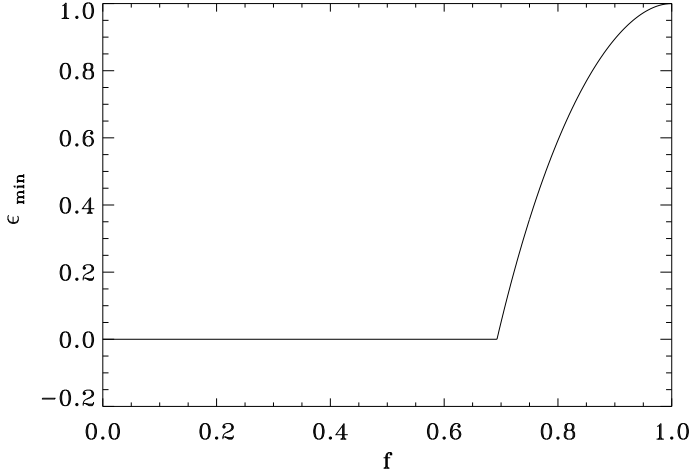,width=10.0cm}
\caption{Minimum value  of the $\epsilon$  parameter in order  to keep
  real the eigenvalues $\tilde{\lambda}^+$ and $\tilde{\lambda}^-$
  plotted as a function of the reduce flux $f=F_r/cE_r$.}
\label{fepsmin}
\end{figure}

It is  worth noticing that in  order to keep  the previous eigenvalues
real  one must have  $(\epsilon \chi')^2  +4(\chi -  f\chi') \geq  0$. 
Therefore, for a given value of  $f$ there is a minimum possible value
($\epsilon_{min}$) for the $\epsilon$ parameter. $(f\chi'-\chi)$ is an
increasing  function of  $f$ which  vanishes for  $f=f_0=2\sqrt{3}/5$. 
Thus, if  $f<f_0$ there is not  constraint on the  value of $\epsilon$
(i.e.  $\epsilon_{min}=0$) but if $f>f_0$ there is a strictly positive
minimum value  for $\epsilon$.   The function $\epsilon_{min}(f)  $ is
plotted  on figure (\ref{fepsmin}).   These limitations  on $\epsilon$
eventhough they  have a mathematical origin,  are physically justified
since they imply that when the radiative flux is high one can not tend
toward the  diffusion scheme.  Futhermore,  we can deduce  from system
(\ref{sysm}) and from  the fact that the $M_1$  model has an intrinsic
flux limitation that $\tilde{f} = \ft_r/cE_r \leq 1$.  It is therefore
important to always keep $\epsilon > f$ in order to compute correctly
the radiative flux. \\

In this  case, we need also to  find as in the  previous subsection an
estimation of  the largest and smallest physical  signal velocities in
the exact  solution to the  Riemann problem. This  estimation involves
the eigenvalues  of the Roe's  linearization for modified  $M_1$ model
\eqref{sysm}. The  Roe's matrix $\mathbf{J_{\ip}}$ for  this model can
be written under the following form:
\begin{equation*}    
  \mathbf{J_{\ip}} = 
  \begin{pmatrix} 
    0 & 1 \\
    (\chi_a - \chi'_m f_a)& \epsilon \chi'_m
  \end{pmatrix},
\end{equation*}
where   we  have   used  the   same  notation   as  in   the  previous
section.  The  eigenvalues $\tilde{\lambda}^{\pm}_{\ip}$  of  this
matrix are:
$$
\label{vp2222}
\tilde{\lambda}^\pm_{\ip} = \frac{\left(\epsilon \chi'_m \pm \sqrt{(\epsilon
      \chi'_m)^2 +4(\chi_a - f_a \chi'_m) } \right)}{2}.
$$

As in the previous case, one can give an estimation of the largest and
smallest physical signal velocities:
\begin{equation}
\tilde{a}^+ = max(\tilde{\lambda}^+_{i+1},\tilde{\lambda}^+_{\ip},0), \ \ \
\tilde{a}^- = min(\tilde{\lambda}^-_i,\tilde{\lambda}^-_{\ip},0).
\end{equation}

Using the HLLE solver gives the following expressions for the flux:
\*
\begin{equation}
\ft_{r,\ip} = \frac{\tilde{a}^+\ft_{r,i} - \tilde{a}^- \ft_{r,i+1}
}{\tilde{a}^+-\tilde{a}^-}  
            + \frac{\tilde{a}^+\tilde{a}^-}{\tilde{a}^+ - \tilde{a}^-}
            (E_{r,i+1}-E_{r,i}) 
\end{equation}

\begin{equation}
P_{r,\ip} = \frac{\tilde{a}^+ P_{r,i} - \tilde{a}^- P_{r,i+1} }
{\tilde{a}^+-\tilde{a}^-} 
          + \frac{\tilde{a}^+\tilde{a}^-}{\tilde{a}^+ - \tilde{a}^-} 
(\ft_{r,i+1}-\ft_{r,i})
\end{equation}

Rewritting the system  in the original variables, we  get the following
discretisation for the $M_1$ model:
\begin{align}
\frac{1}{\dt} (E^{n+1}_r - E^n_r)_i &+ \frac{1}{\dr} (F_{r,\ip} - F_{r,\im})^\np = \sigma c (a_r T^4 - E_r)_i^\np\\
\frac{1}{\dt} (F^{n+1}_r - F^n_r)_i &+ \frac{c^2}{\dr} (\hat{P}_{r,\ip} - \hat{P}_{r,\im})^\np = -\sigma c F^\np_{r,i}
\end{align}
\noindent where we have defined: 
$$\hat{P}_{r,\ip} = \epsilon^2 P_{r,\ip} + (1-\epsilon^2)(P_{r,i+1} + P_{r,i})/2
\;\;\;\; \mbox{  and  }\;\;\;\; 
F_{r,i\pm\frac{1}{2}} = c \epsilon \ft_{r,i\pm\frac{1}{2}}$$

The method  is easily extended  to the radiation subsystem  written in
the comoving frame but where the nonconservative terms are dropped and
with a moving grid. The sytem then writes 
\begin{align}
\partial_t E_r & +\; \partial_x((u-u_g)E_r) +\;\partial_x  F_r = 
\sigma c (a_r T_m^4 - E_r),
\label{equ1_1d_comobile}\\ 
\partial_t F_r & + \partial_x((u-u_g)F_r) + c^2\partial_x \mathbf{P}_r =
 -\sigma c Fr, \label{equ2_1d_comobile}
\end{align}
The non-conservative  terms are dropped  for the present  analysis but
they must naturally  be taken into account as will  be detailed in the
next section. The  previous systeme can be written  in a dimensionless
way
\begin{align}
  \partial_{\tilde{t}} E_r & + \partial_{\tilde{x}} (\frac{\tilde{u}}{\epsilon} E_r + \tilde{F_r})  
  = \frac{(a T_m^4 - E_r)}{\epsilon^2}, \\
  \partial_{\tilde{t}} \tilde{F_r} & + \frac{1}{\epsilon^2}\partial_{\tilde{x}} \chi E_r
  + \frac{1}{\epsilon} \partial_{\tilde{x}} (\tilde{u} \tilde{F_r}) 
  = -\frac{1}{\epsilon^2} \tilde{F_r}, 
\end{align}
where $$\tilde{u} = \frac{u - u_g}{c} \cdot$$

As for the system (\ref{equ1_mod})-(\ref{equ2_mod}) we transform the
previous system in a non-stiff hyperbolic system with stiff source
term
\begin{align}
  \partial_{\tilde{t}} E_r + \partial_{\tilde{x}} (\tilde{u} E_r + \tilde{F_r})  
  = & \; \; \; \; \frac{1}{\epsilon^2} \left ( (a T_m^4 - E_r) 
    - \epsilon (1 - \epsilon) \partial_{\tilde{x}} (\tilde{u} E_r) \right ), 
  \\
  \partial_{\tilde{t}} \tilde{F_r}  + \partial_{\tilde{x}} (\chi E_r +
  \tilde{u} \tilde{F_r}) 
  = & -\frac{1}{\epsilon^2}
  \left (\tilde{F}+(1-\epsilon^2)\partial_{\tilde{x}} \chi E_r 
    + \epsilon (1 - \epsilon) \partial_{\tilde{x}} (\tilde{u}
    \tilde{F_r}) \right), 
\end{align}
whose eigenvalues are
\[
\label{vp2bis}
\tilde{\lambda}^\pm =  \frac{\left(\epsilon \chi'\pm \sqrt{(\epsilon \chi')^2 
+4(\chi - f\chi') } \right)}{2}+\tilde{u}.
\]
The same analysis as above leads to the following scheme
\begin{align}
\frac{1}{\dt} (E^{n+1}_r - E^n_r)_i &+ \frac{1}{\dr} (F_{r,\ip}^* -
F_{r,\im}^*)^\np = \sigma c (a_r T^4 - E_r)_i^\np\\ 
\frac{1}{\dt} (F^{n+1}_r - F^n_r)_i &+ \frac{c^2}{\dr}
({P}_{r,\ip}^* - {P}_{r,\im}^*)^\np = -\sigma c F^\np_{r,i} 
\end{align}
where 
\begin{eqnarray}
F_{r,\ip}^* &=&F_{r,\ip}+ \epsilon \ \frac{\tilde{a}^+(u_{\ip}-u_g)E_i-\tilde{a}^
-(u_{\ip}-u_g)E_{i+1}}{\tilde{a}^+-\tilde{a}^-}\\
&+& \frac{(1-\epsilon)}{2} \ (u_{\ip}-u_g) (E_{r,i+1} + E_{r,i})), \notag \\
P_{r,\ip}^* &=& \epsilon^2 P_{r,\ip} + \frac{(1-\epsilon^2)}{2} \ (P_{r,i+1} + P_{r,i})\\
& + & \frac{\epsilon}{c^2} \ \frac{\tilde{a}^+((u_{\ip}-u_g))F_{r,i}-\tilde{a}^-
((u_{\ip}-u_g))F_{r,i+1}}{\tilde{a}^+-\tilde{a}^-} \notag \\
&+& \frac{(1-\epsilon)}{2 c^2}\ (u_{\ip}-u_g) (F_{r,i+1} + F_{r,i}), \notag
\end{eqnarray}
and
\begin{equation}
\tilde{a}^+ = max(\tilde{\lambda}^+_{i+1},\tilde{\lambda}^+_{\ip},0), \ \ \
\tilde{a}^- = min(\tilde{\lambda}^-_i,\tilde{\lambda}^-_{\ip},0).
\end{equation}

In the  previous discretization of  the radiation subsystem,  one need
the value of $\epsilon$  at each interface. In practical computations,
where  the  opacity gradient  can  be large,  we  compute  a value  of
$\epsilon$ in  each cell $\epsilon  = 1/\sigma\Delta x$. The  value of
$\epsilon$ at the  interface is then taken to be  the maximum value in
the two  adjacent cells.  As  we will see  this method has  given very
satisfactory results  even in  the case of  radiative shock  with very
strong opacity gradient.\\

Let      us      also       remark      that      if      we      take
$\tilde{a}^{\pm}=\tilde{\lambda}_{\ip}^{\pm}$,  then  the HLLE  solver
coincides with the Roe solver for the radiation subsystem. Thus, if we
use for the hydrodynamic subsystem a Roe solver, then the above scheme
is exactly the Roe scheme for the radiation-hydrodynamics system where
the non conservative coupling terms are dropped.  Nevertheless using
HLLE solver for the radiation insures positivity of the scheme. \\

\section{Actual implementation of the method}

We present in this section how the full radiation-hydrodynamics system
is discretized  and integrated.  As  we mentionned before,  when using
system  (\ref{hydro1}-\ref{rayo2}) the  treatment of  hydrodynamics is
independant of  the radiation  (apart form the  source terms)  and has
been  described in section  (\ref{section_hydro}).  For  the radiative
transfer in spherical  coordinates the equations to be  solved are the
following:

\begin{equation}
\label{equray}
\left\{
\begin{array}{ll}
\partial_t (E_r S) & + \;\;\;\partial_r (F_r S) +  \partial_r ((u-u_g)E_rS)
+ P_rS \nabla . u \\ & = \;\; \;\kappa\rho c S (a_r T^4 - Er) + 
\frac{u}{r}S(3P_r -E_r)
\\ \\
\partial_t (F_r S) & + \;\;\;c^2\partial_r (P_r S)+ \partial_r ((u-u_g)F_rS)
+ F_r S \partial_r u \\ & =\; - \kappa\rho c F_rS + 
c^2 P_r\partial_r S  - \frac{(3P_r -E_r)}{r}c^2 S  
\end{array}
\right.
\end{equation}

The   left-hand   sides   of   the  previous   system   contains   the
$M_1$-radiation  subsystem  and coupling  terms  due  to the  comoving
frame.  We treat the  conservative terms using the Godunov-type method
presented  in  the  previous  section.  The  nonconservative  coupling
terms,  which,  as  we  have   seen  previously,  do  not  modify  the
eigenvalues of  the physical system, are discretized  in the following
way: the  velocity divergence  is discretized using  the value  of the
velocity at the  interface given by the hydrodynamic  solver.  The two
terms of the right-hand-side are local and are computed by integration
over the volume of the cell.  All the terms of the second equation are
discretized in the  same maner.  And finally, all  terms (given by the
hydrodynamic  or  radiation solver,  comoving  and  source terms)  are
computed at the  same time so that all equations  are fully coupled in
an implicit scheme.  This is summerized in the following tables.

\begin{equation*}
\begin{array}{|c|c|c|}
\hline
\kappa\rho c S (a_r T^4 - E_r) & P_rS \nabla . u & \frac{u}{r}S(3P_r -E_r)\\
\hline
\left(\kappa\rho c (a_r T^4 - E_r)\Delta v\right)_i^{\np}    
&   (P_r)_i^{\np} \left(uS\right)_\gr^{\np}          &  \left(\frac{u}{r}(3P_r -E_r)\Delta v\right)_i^{\np}     \\
\hline
\end{array}
\end{equation*}

\begin{equation*}
\begin{array}{|c|c|c|c|}
\hline
\kappa\rho c F_rS & c^2 P_r\partial_r S  & \frac{(3P_r -E_r)}{r}c^2S &
F_r S \partial_r u \\ 
\hline
\left(\kappa\rho c F_r\Delta v\right)_i^{\np} & \left(c^2 P_r S \right)_i^{\np} & 
\left(\frac{(3P_r -E_r)}{r}c^2\Delta v\right)_i^{\np} &  (F_r)_i^{\np} \left(uS\right)_\gr^{\np} \\
\hline
\end{array}
\end{equation*}

In order to gain in accuracy, we actually use a classical second order
MUSCL  method (\cite{vl}) to compute  the flux.  For the hydrodynamics
{\it minmod}  slopes  are computed  for  $\rho, u$ and  $P$.   For the
radiative transfer, slopes are computed for $E_r$  and $f$ in order to
always satisfy the constraint $ |f| \le 1$. These slopes are then used
to  extrapolate  the  value of  the   different fields near   the cell
interface.  These extrapolated values are  then used as inputs for the
Riemann solvers and for the upwind flux.
 
Finally, as we will show in section 5.4, the CFL condition given by an
explicit  treatment  of  hydrodynamics  is  too  much  small  in  some
applications  and  a full  implicit  treatment  is  needed.  Then  the
non-linear       system       of      equations       \eqref{equrada},
\eqref{equdens}-\eqref{equener}    and   \eqref{equray}    is   solved
iterativelly using  a Raphson-Newton  method.  The jacobian  matrix is
computed  using numerical  derivatives. We  use for  the implicitation
parameter $\alpha = 0.55$.

\section{Test cases}

We   present    in   this   section   some   test    cases   for   the
radiation-hydrodynamic  scheme presented  above.  We  start  with some
simple shock-tube  problems and end with  the collapse of  a gas cloud
leading to  a stellar-like structure.  All the  shock-tube problems are
performed using  a constant and uniform resolution  (i.e. eulerian grid)
in   order   to   place    emphasis   on   the   properties   of   the
radiation-hydrodynamics solver.

\subsection{Radiative front}
\begin{figure}[httb]
\begin{center}
\psfig{file=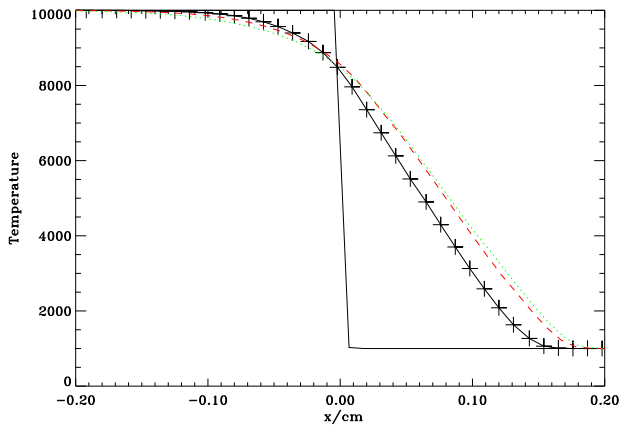 ,height=6.0cm}
\caption{\it  The  radiative  temperature  in the  initial  conditions
(solid  line),  after a  propagation  of  1cm  (ie 100  cells)  (black
cross), 2cm  (dashed line) and  3cm (dotted line). Each  curve is
shifted along the x-axis of  $-c\dt$. Which means that since the front
propagate  at the  speed of  light  the discontinuity  should stay  at
$x=0$.}
\label{fig_front}
\end{center}
\end{figure}

We first  look at the propagation  of a radiative front  in vacuum. In
the initial  conditions the half-space $x<0$ is  filled with radiation
with $T_r  = 10^4$ and $f=1$.   For the half-space $x>0$  we have $T_r
=10^3$ and $f=1/3$.  The front  should propagate at the speed of light
and   remain   sharp.    The   results   are   presented   on   figure
(\ref{fig_front}).  We first see that,  thanks to the $M_1$ model, the
front propagates at  the proper speed. After a  propagation of 1cm (ie
100  cells)  we see  that  the  front  has suffered  from  significant
diffusion and is spread among  about 20 cells. But, when it propagates
further the  spatial extension  of the front  remains roughly  stable. 
This stability of the front is due to the fact that for $f=1$ the wave
speed  is equal  to  $c$  while for  $f=1/3$  it is  equal  to $c/3$.  
Therefore,  contrary to  a contact  discontinuity, there  is  a slight
compression of this radiative front.

\setlength{\unitlength}{1.0cm}

We  now  test our  scheme  with  radiative  shocks playing  particular
attention to the  stability of the solution for  different sampling of
the photon mean free path.

\subsection{Radiative shock dominated by hydrodynamical pressure}

We  examine in this  section a stationary strong  shock  in a gas with
unit molecular weight and a constant  opacity of $\kappa = 4.0\;10^4\;
cm^ 2/g$.  The upstream quantities for the gas are given by:
\begin{equation}
\rho = 1.2 \; 10^{-3}  \;g/cm^3 \;\;\;\;\;\; T = 10  \;^o K  \;\;\;\;\;\; u= 200 \;km/s 
  \;\;\;\;\;\;T_r = T  \;\;\;\;\;\; F_r = 0.
\end{equation}

As the upstream material is  extremely opaque, any radiation  incoming
from   the  radiative front   will be   completely reabsorbed at  some
distance.  Far from this   area,  the flow  becomes  homogeneous,  the
radiative  flux  vanishes ($F_r =   0$),  the radiation  and the fluid
temperatures are in equilibrium ($T_r = T$).

The downstream quantities are chosen so that the Rankine-Hugoniot jump
conditions are satisfied for the mass \eqref{hydro1}, total momentum
\eqref{totmom} and total energy \eqref{toten}. Let us notice that in
that case the   Rankine-Hugoniot  jump  conditions  for  the  complete
coupled fluid  radiative system  (\ref{hydro1} - \ref{rayo2})  are not
satisfied.  As we introduce only three   constraints on the downstream
state, we need two extra relations to know this state (see Mihalas
\cite{mihalas}).  The  downstream material  is also  extremely opaque,
hence, we still have $F_r = 0$ and $T_r =  T$.  In fact, the gas given
by the downstream or  upstream states act as  a  real non  perfect gas
where the material pressure $P$ is replaced by the total pressure $P +
P_r$ and where the material energy $E$ is replaced by the total energy
$E + E_r$. In this test, the equation of state is totally dominated by
the gaz ($\gamma = 5/3$) and  radiation is only significant for energy
transport.

The  photon mean  free  path  in  the  upstream material is   given by
$\lambda   = 1/\kappa\rho  \simeq    0.02  cm$.   The results  of  the
simulation for a   grid with $\dx  =  0.5\lambda$ are given in  figure
(\ref{fig_choc1}). The hydrodynamical shock \*is stable and is sampled
over 2 cells and there is  a large region in  front of the shock where
the gas is  preheated by radiation. In this  thermal precursor the gas
is heated and slowed down by the energy released in the shock. We have
looked at the behavior of this  precursor under changes in resolution.
We  can see (figure (\ref{fig_prec5}-A)) that  even when the mean free
path of photon  is not sampled,  the thermal precusor keeps the proper
spatial extension.   If the numerical  scheme is not modified  to take
into  account the diffusion  limit, the extension  of the precursor is
much too   large,  as   was expected  from     the analysis  made   in
(\ref{modif}). If  the spatial sampling  of the \* simulation is large
compared  to the  length of  the   precursor, the temperature  jump is
captured  over \* 2  or 3  cells which is   roughly as accurate as the
treatment of a shock.

\setlength{\unitlength}{1.0cm}
\begin{figure}[httb]
\begin{center}
\begin{picture}(12,12)
\put(0.0,5.5){
\psfig{file=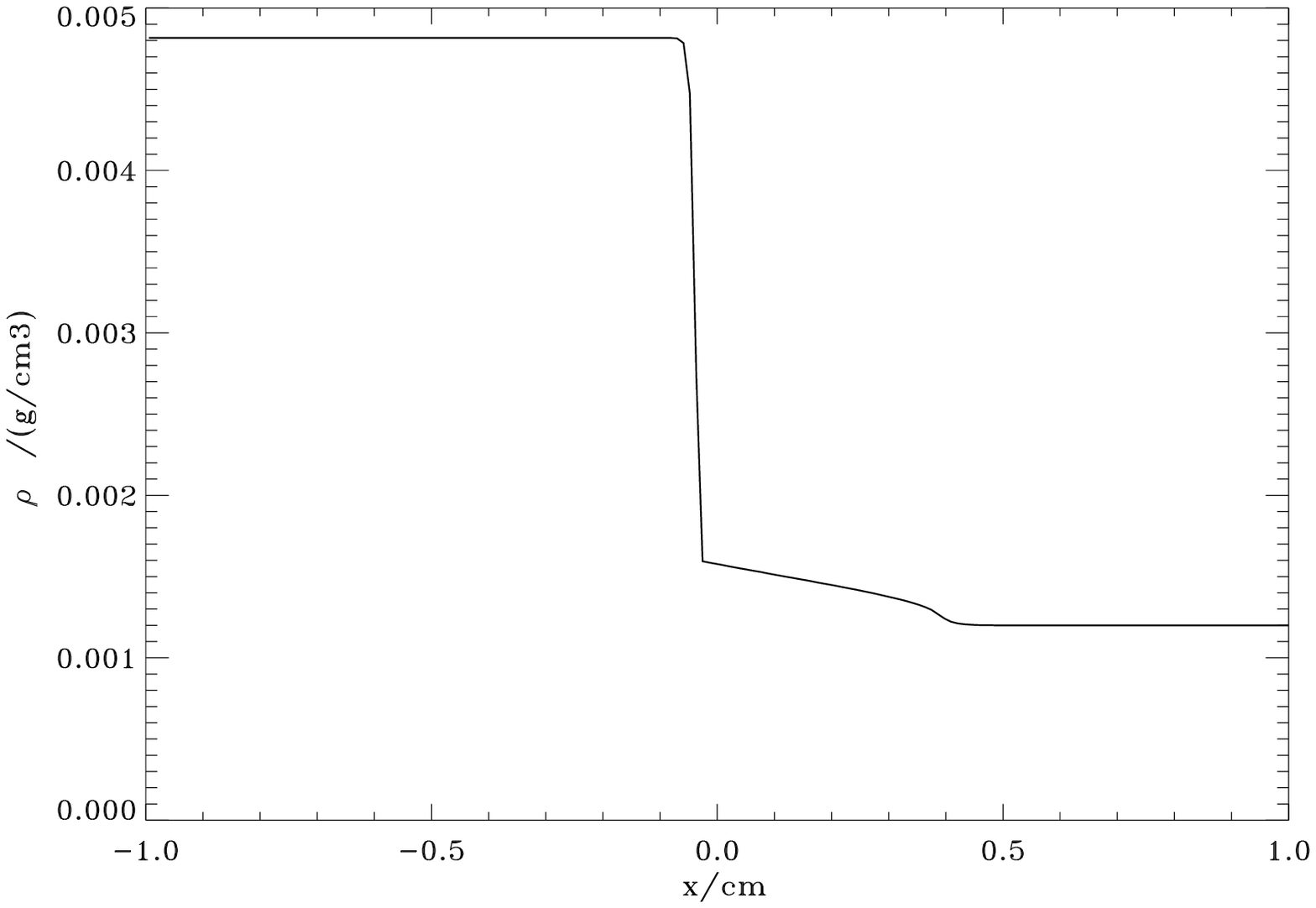 ,height=5.0cm,width=5.0cm}}
\put(5.5,5.5){
\psfig{file=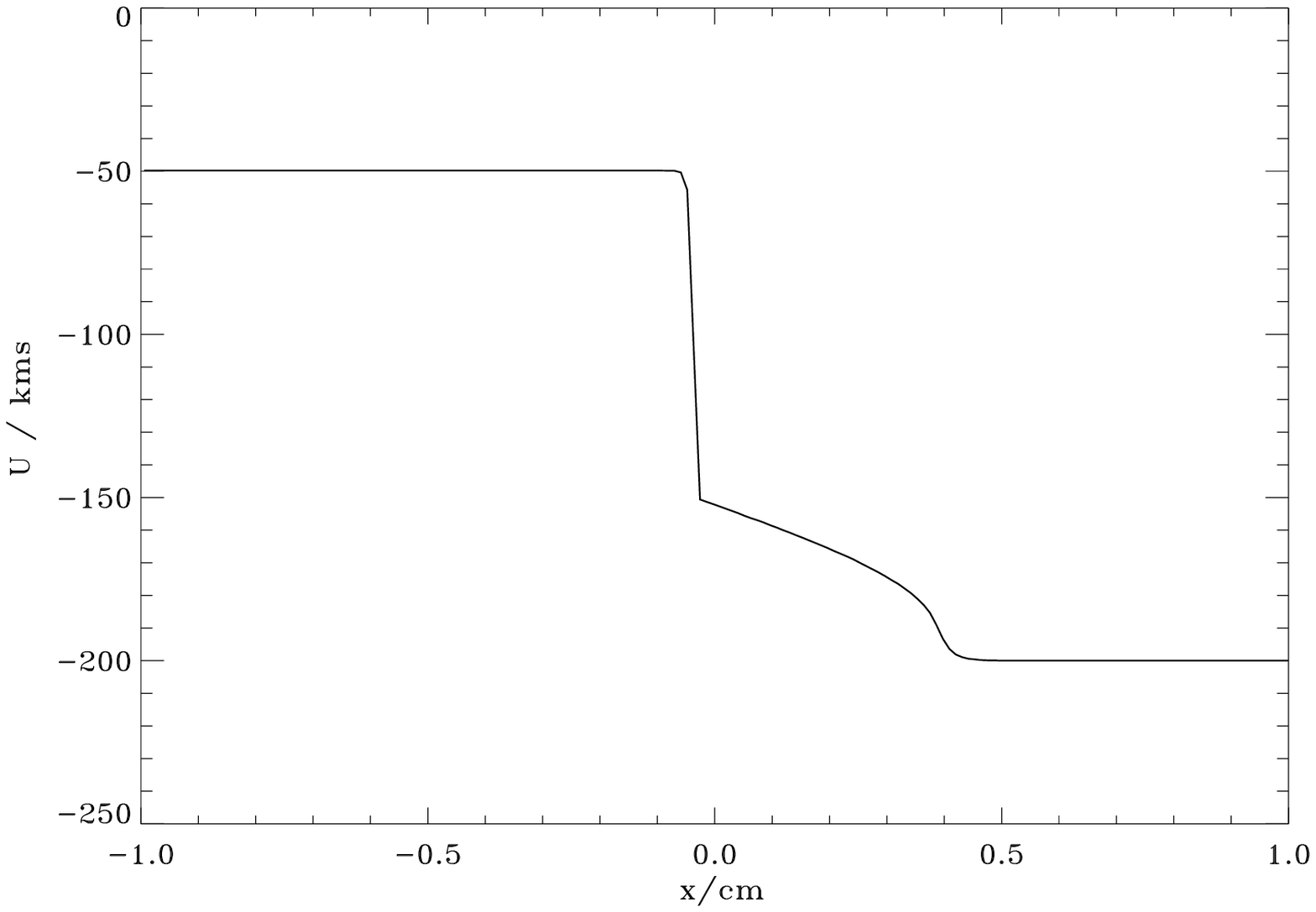   ,height=5.0cm,width=5.0cm}}
\put(0.0,0.0){
\psfig{file=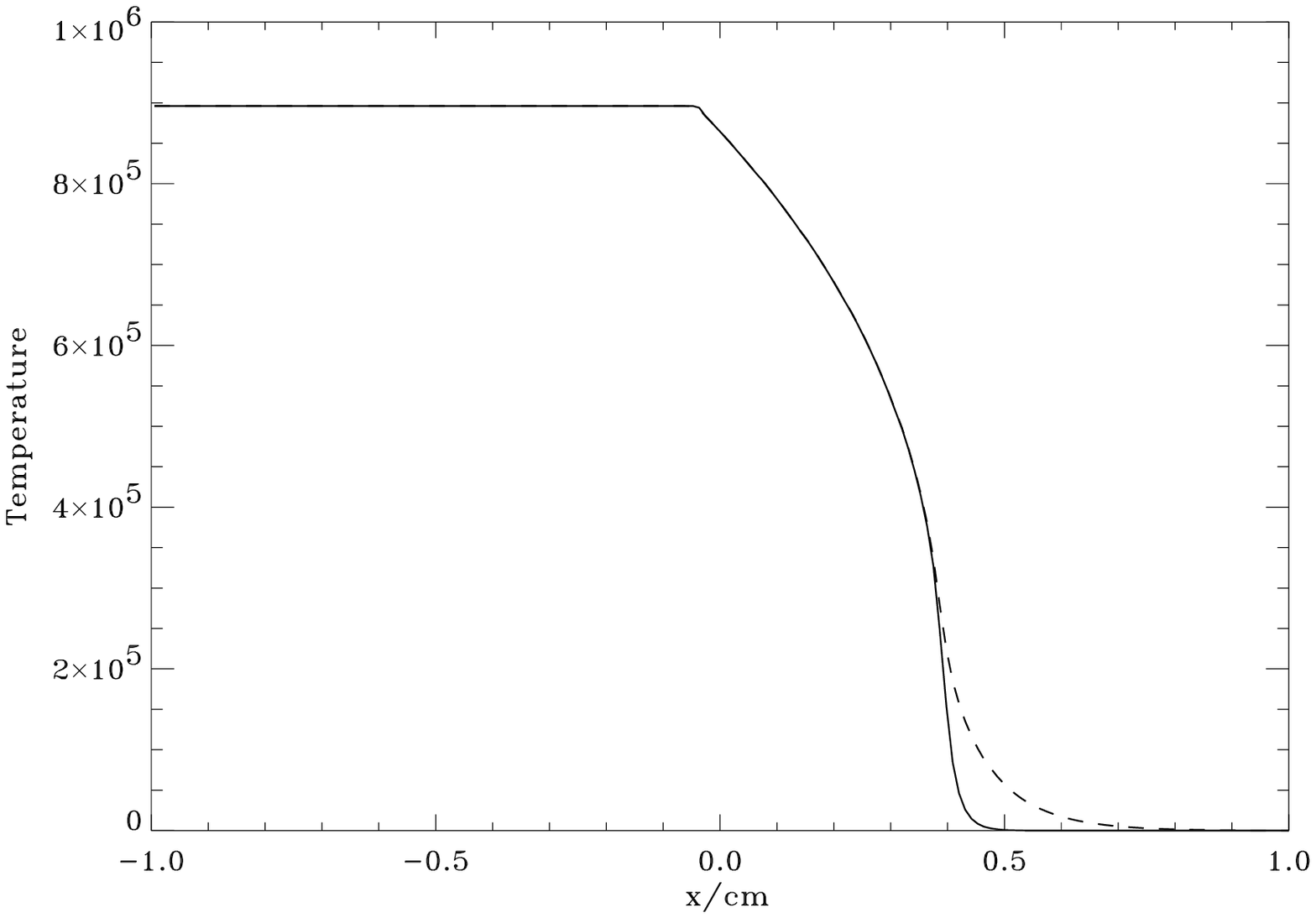,height=5.0cm,width=5.0cm}}
\put(5.5,0.0){
\psfig{file=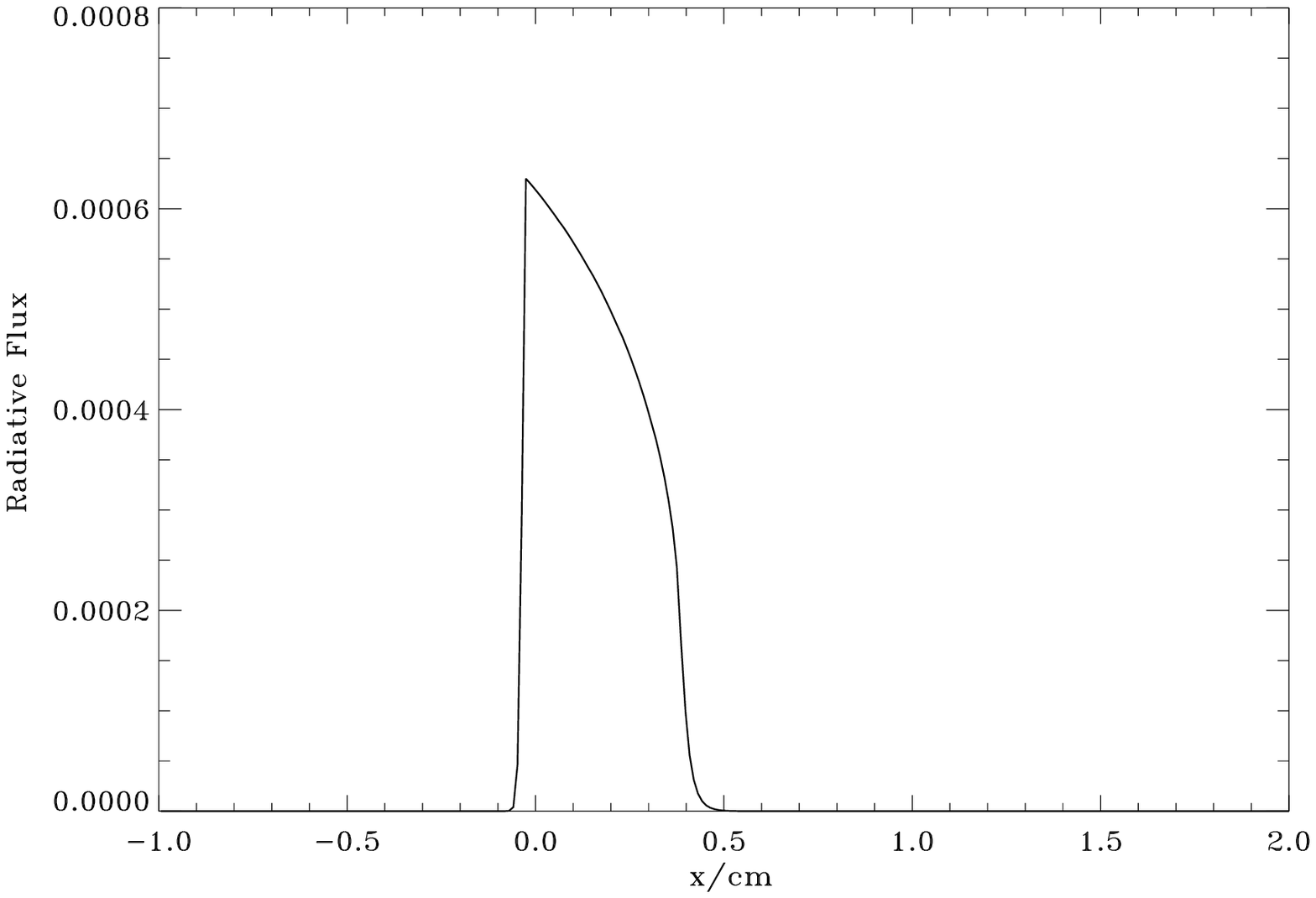  ,height=5.0cm,width=5.0cm}}
\put(1.5,6.5){ \bf A }
\put(6.8,6.5){ \bf B }
\put(1.5,1.0){ \bf C }
\put(6.8,1.0){ \bf D }
\end{picture}
\end{center}
\caption{\it Profiles  of density (A), velocity (B),  gas (solid line)
and radiation  (dashed line) temperatures  (C) and radiative  flux (D)
for a strong shock with negligible radiative energy density.}
\label{fig_choc1}
\end{figure}

\setlength{\unitlength}{1.0cm}
\begin{figure}[httb]
\begin{picture}(16,6)
\put(-0.5,0.0){
\psfig{file=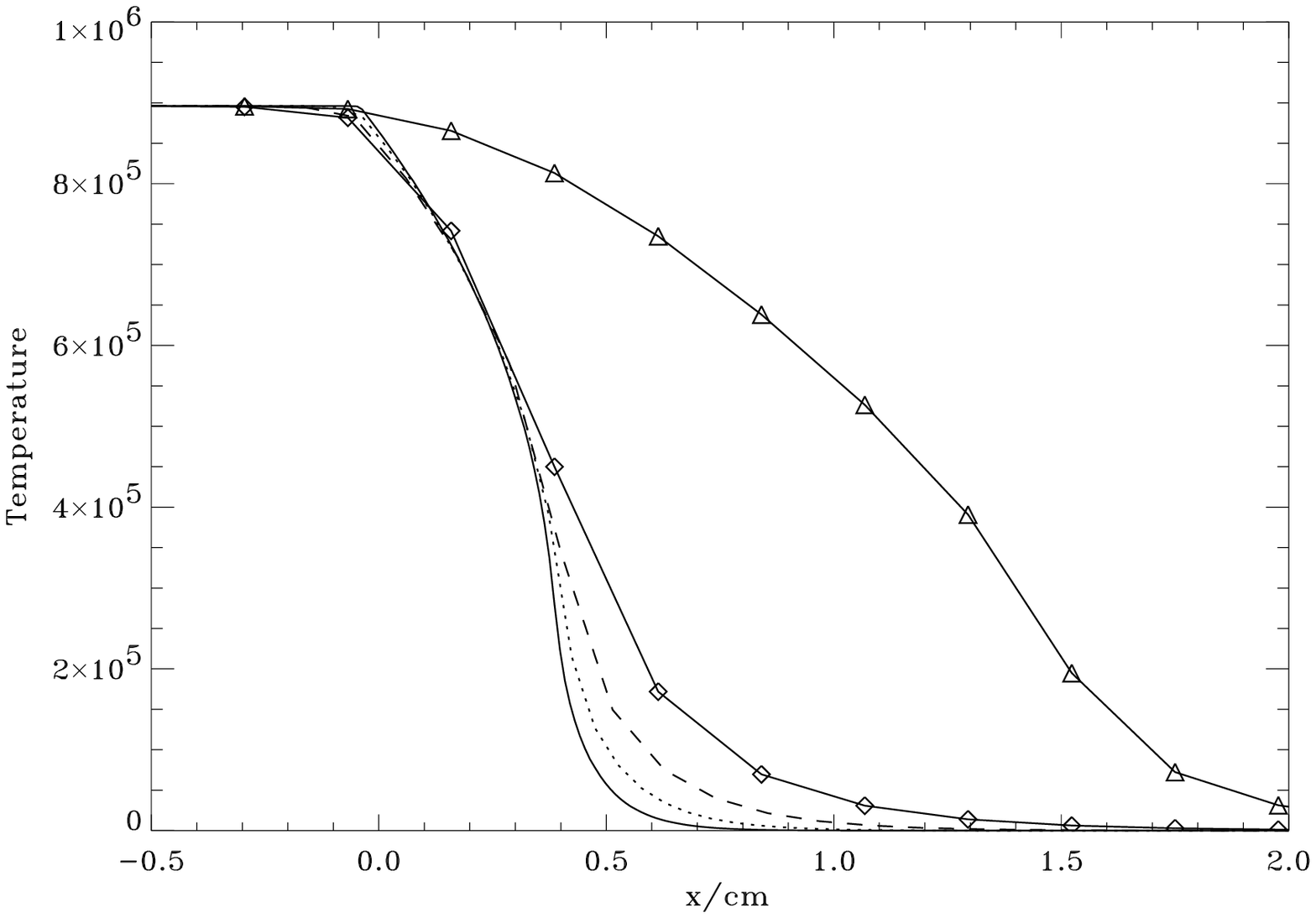 ,height=5.0cm}}
\put(7.0,0.0){
\psfig{file=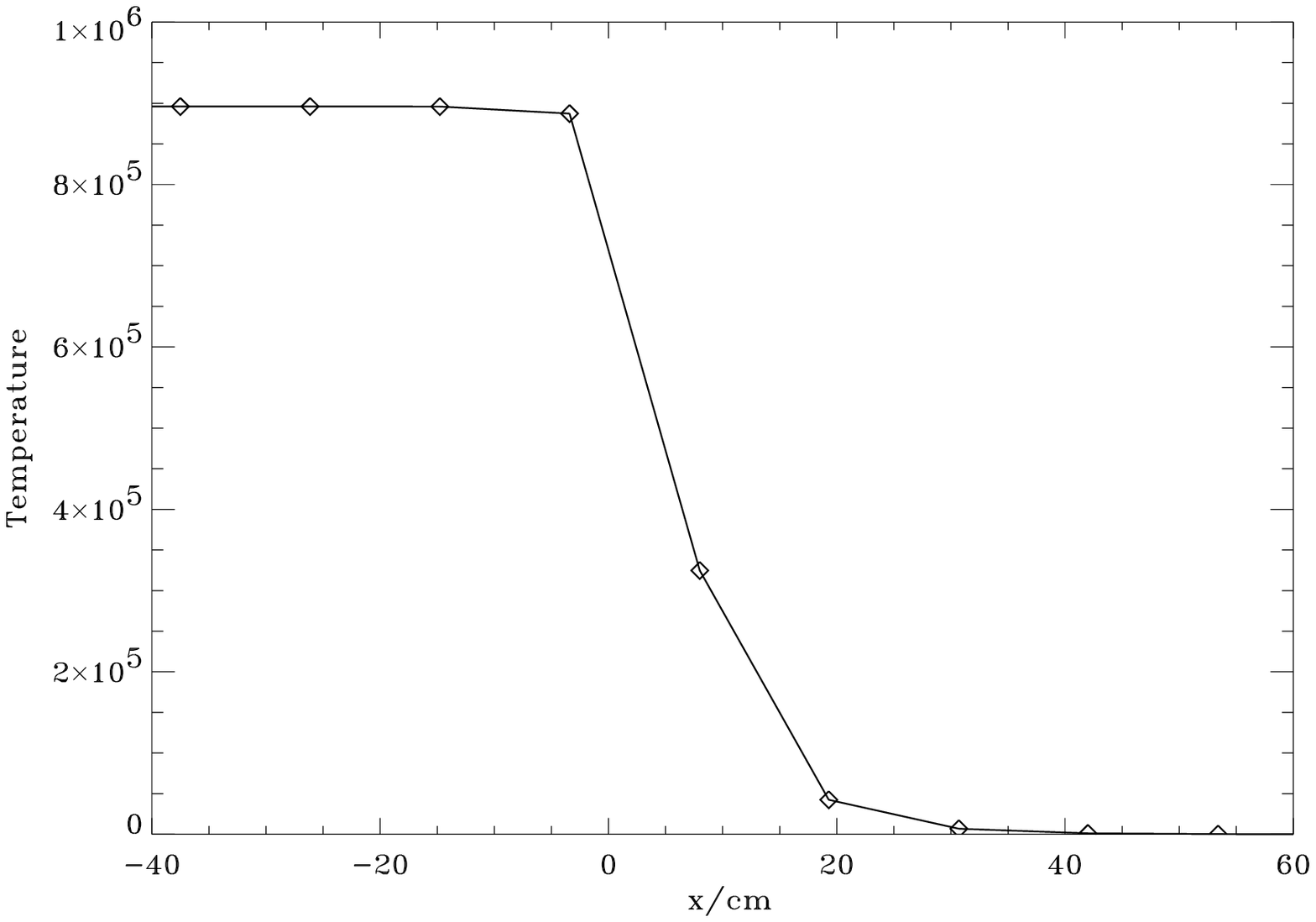 ,height=5.0cm}}
\put(1.0,1.0){ \bf A }
\put(8.3,1.0){ \bf B }
\end{picture}
\caption{\it  Profiles  of  gas   temperature  in  the  upstream  flow
  preheated by the  radiation.  In plot A the grid  spacing was $\dx =
  0.5\lambda$ (solid  line), $2.5\lambda$ (dotted  line), $5.5\lambda$
  (dashed  line), $11 \lambda$  (solid line  with squares).  The solid
  line  with  triangle is  the  results of  a  simuation  with $\dx  =
  11\lambda$  but  where  $\epsilon$   was  kept  to  unity.  For  the
  simulation of plot B the grid spacing was $550 \lambda$, the thermal
  precursor  is now  sampled on  2-3 grid  points which  is  almost as
  accurate as the treatment of a shock.}
\label{fig_prec5}
\end{figure}

\subsection{Shock dominated by radiative pressure}

We now  examine the case of  a shock where the  dynamical influence of
radiation is not  limited to energy transport. In  this test case, the
energy in the downstream region  is dominated by the radiative energy. 
The upstream conditions are:

\begin{equation}
\nonumber
\rho =  10^{-3}  \;g/cm^3 \;\;\;\; T = 10^5  \;^o K  \;\;\;\; u = 1000 
 \;km/s \;\;\;\; T_r = T  \;\;\;\; F_r = 0.
\end{equation}

\noindent 
and  the  downstream  conditions  are given  by  the  Rankine-Hugoniot
relations as detailed  in the above section.  As  in the previous test
we assume for the radiation that downstream states are at equilibrium,
(i.e. $E_r = a_r T^4$ and $F_r = 0$).

\begin{figure}[httb]
\begin{center}
\psfig{file=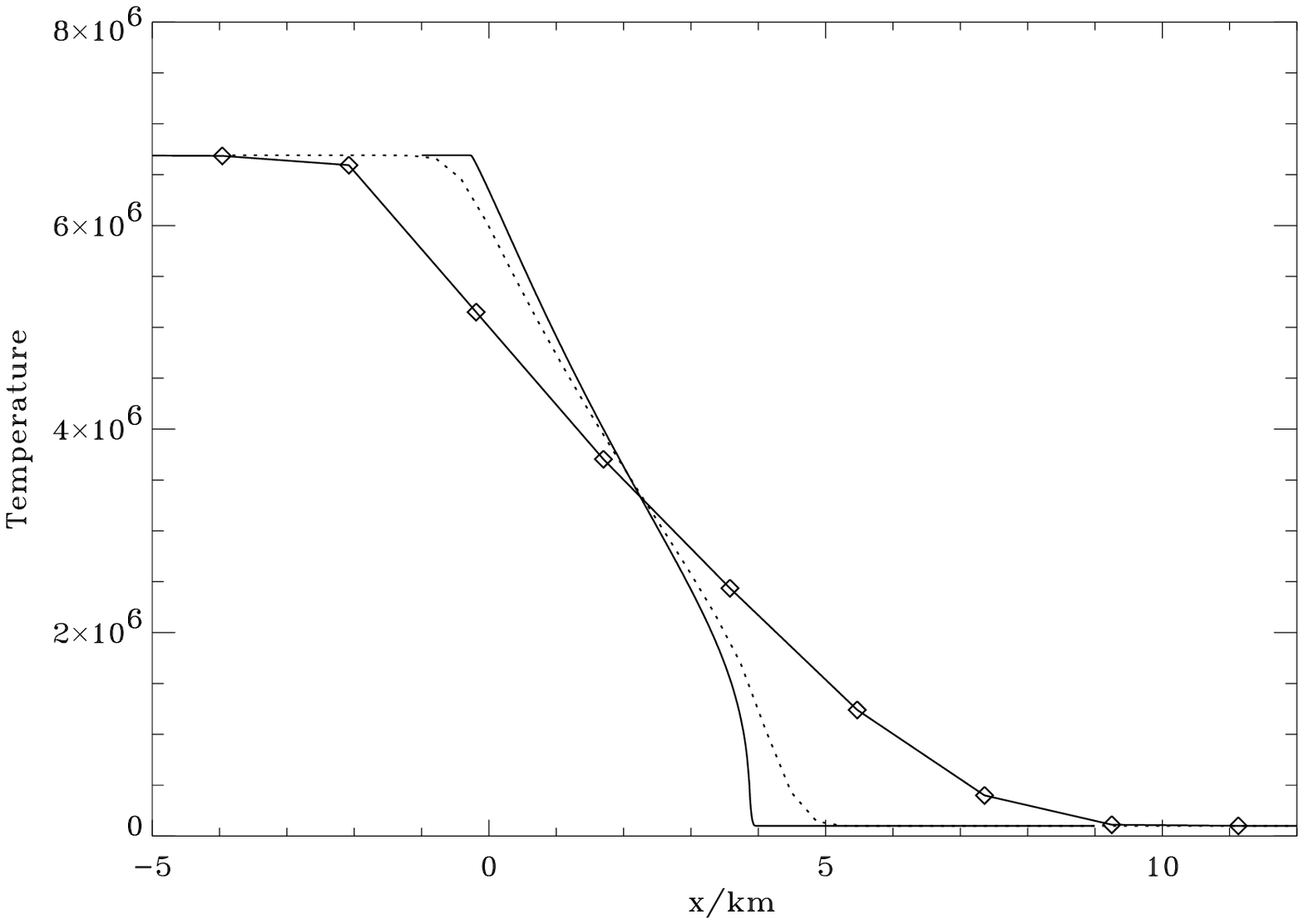 ,height=7.0cm}
\end{center}
\caption{\it Profiles of temperature in the upstream flow preheated by
the  radiation. The grid  spacing was  $\dx  = 0.5\lambda$
(solid line), $35\lambda$  (dotted line), $200\lambda$ (squares).}
\label{fig_prec1}
\end{figure}

The gas molecular  weight is, as before, set to  unity and the opacity
is $\kappa = 1 \; cm^2/g$.   The photon mean free path in the upstream
region is therefore $\lambda =  1/\kappa\rho = 10^3 \;cm$. The results
of  this test  case for  $\dx =  0.5\lambda$ are  presented  on figure
(\ref{fig_choc2}).  As in the previous  case, the shock is captured in
two cells and there is a large preheated region in front of the shock.
In this region, energy is transfered from the gas to the radiation and
the  radiation energy  density  quickly dominates  as the  temperature
rises. In the  upstream region the equation of  states is dominated by
the gaz ($\gamma = 5/3$) and  in the downstream region it is dominated
by radiation ($\gamma = 4/3$).  In this case the effective $\gamma$ in
the  downstream region is  $4.2/3$.  Eventhough  we do  not explicitly
treat  the total  (gaz  and radiation)  conservation  law this  energy
transfer is treated  correctly. As in the previous  case, the shock is
absolutly  stable  and  the  Rankine-Hugoniot relations  are  strictly
verified.   This  example illustrates  the  fact  that eventhough  our
hyperbolic  analysis decouples  matter  and radiation,  our scheme  is
valid even when  there are large momentum and  energy transfer between
gas and  radiation.  In figure  (\ref{fig_prec1}) we have  plotted the
temperature profile  for different grid resolutions  to illustrate the
proper  behavior of  our scheme  in both  the streaming  and diffusion
regime.

\setlength{\unitlength}{1.0cm}
\begin{figure}[httb]
\begin{center}
\begin{picture}(12,12)
\put(0.0,5.5){
\psfig{file=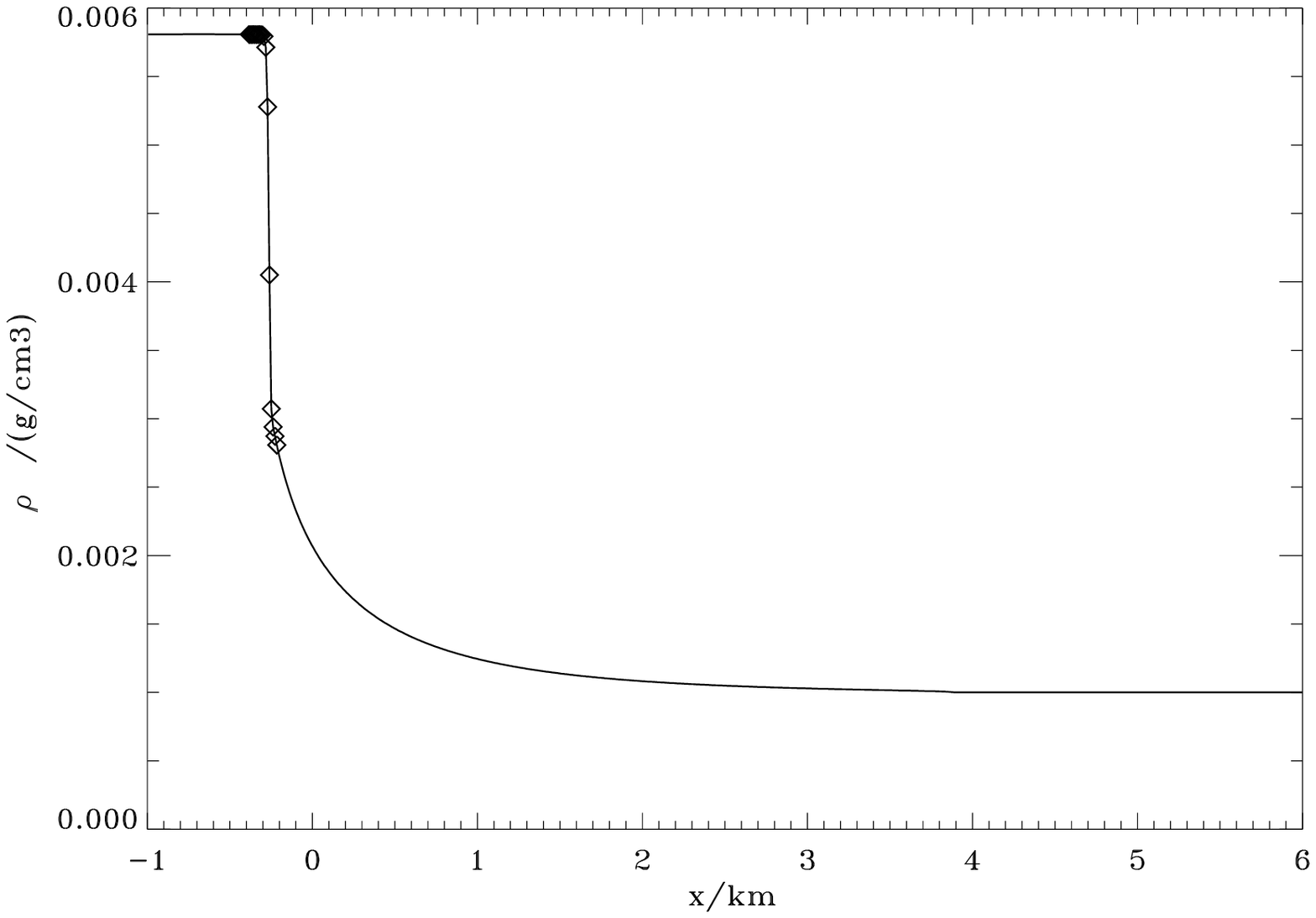 ,height=5.0cm,width=5.0cm}}
\put(5.5,5.5){
\psfig{file=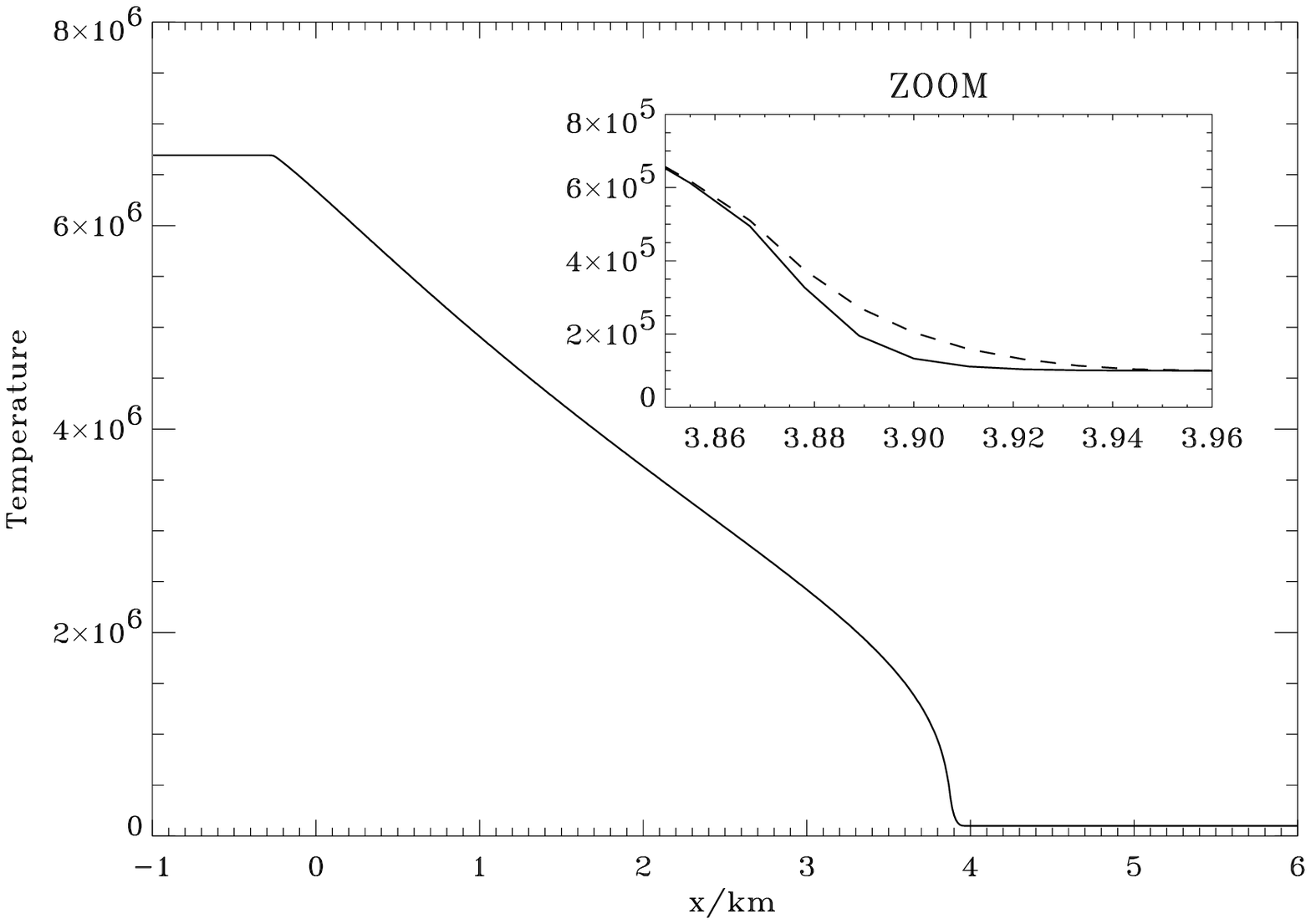   ,height=5.0cm,width=5.0cm}}
\put(0.0,0.0){
\psfig{file=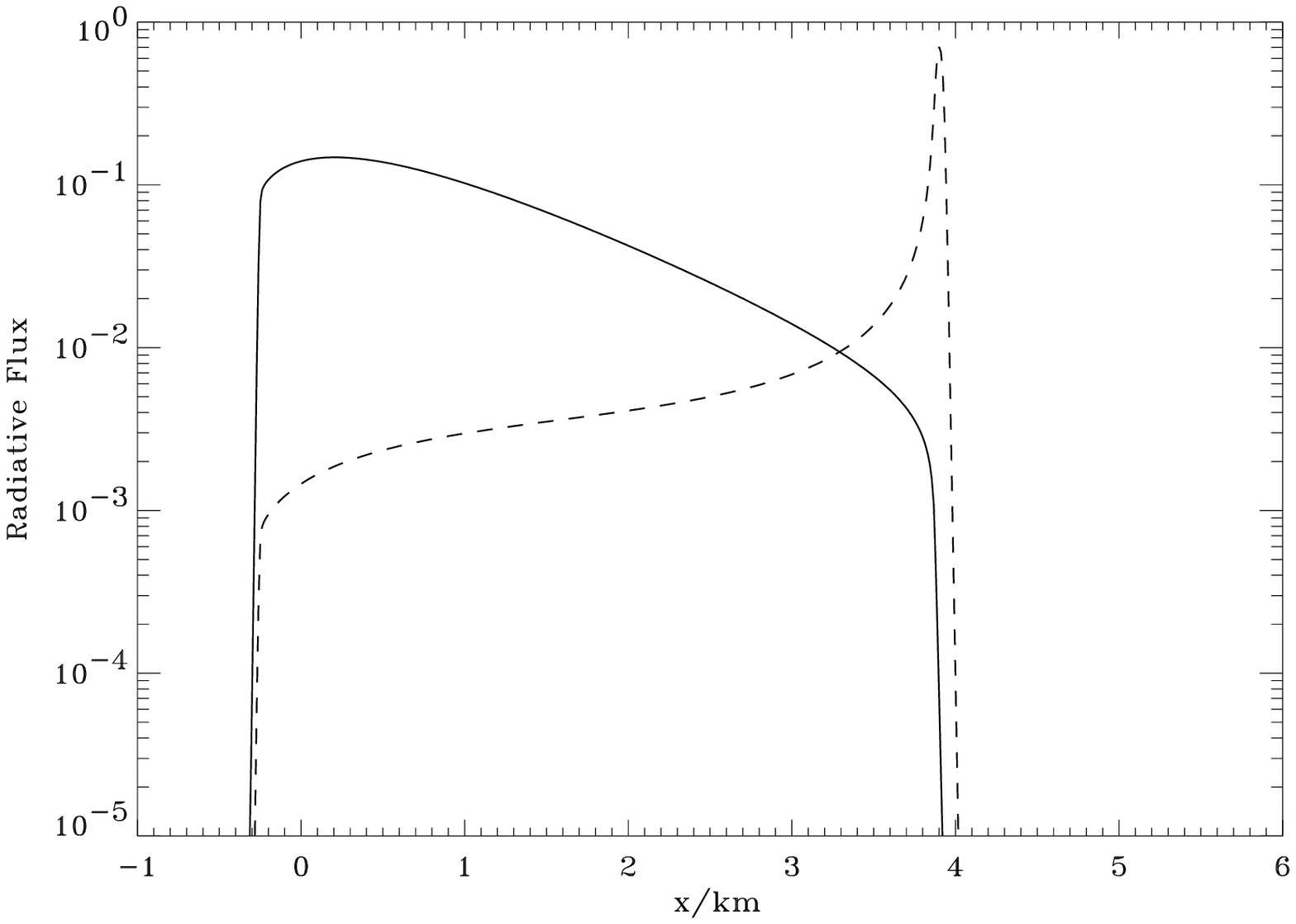,height=5.0cm,width=5.0cm}}
\put(5.5,0.0){
\psfig{file=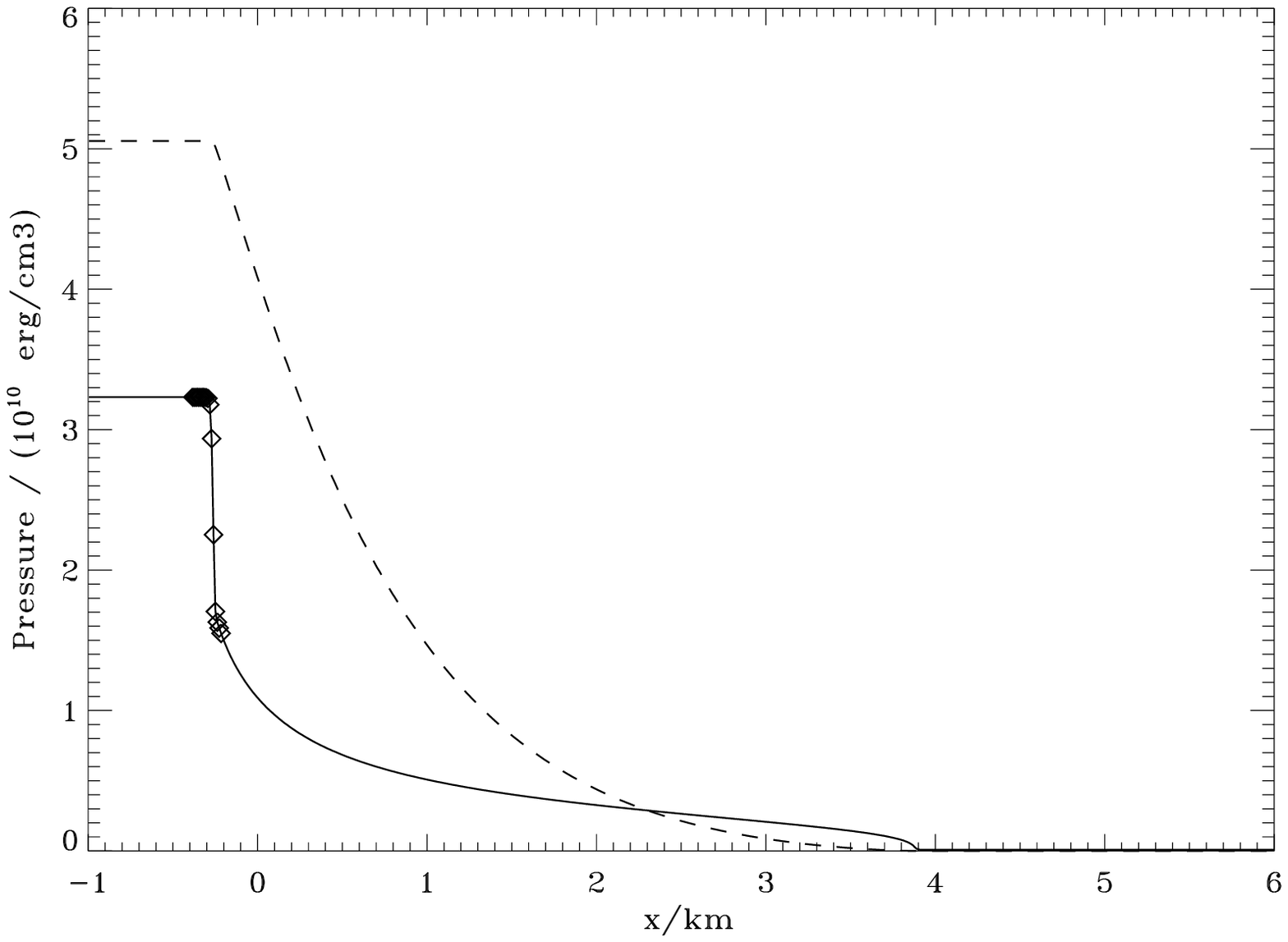  ,height=5.0cm,width=5.0cm}}
\put(1.5,6.5){ \bf A }
\put(9.8,6.5){ \bf B }
\put(1.5,1.0){ \bf C }
\put(9.8,1.0){ \bf D }
\end{picture}
\end{center}
\caption{\it Profiles  of density (A), gas (solid  line) and radiation
(dashed line)  temperatures (B),  radiative flux (in  arbitrary units,
solid line) and reduce flux (dashed line) (C) and gas (solid line) and
radiative (dashed  line) pressures  (D) for a  strong shock  where the
downstream energy is dominated by radiative pressure.  The squares for
the  gas  density and  pressure  show  that  eventhough the  radiative
pressure  dominates   the  hydrodynamical  shock   is  still  properly
treated.}
\label{fig_choc2}
\end{figure}

\subsection{Collapse of a gas cloud}

Star  formation  is a   very    challenging problem,  where  radiation
transport plays a dominant  role.  The modeling of  the formation of a
star introduces a  wide range  of length  and  time scales.  The  mean
density contrast between a collapsing cloud and the star reaching main
sequence  is about   $10^{20}$.  The  initial  cloud  radius  is about
$10^{18}$cm, while  the typical radius   of a solar-like star is  less
than $10^{11}$cm.   The  compactness of  the   star embryo requires  a
spatial   resolution of at least $\Delta   r/r  \simeq 10^{-4}$ in the
envelope outer layers, where the pressure  gradients are the highests.
This means  that the Courant number corresponding  to a dynamical time
scale $t_d =  r/c_s$, where $c_s$ is  the  sound speed, is about  $C =
10^4$.  If the  evolution  of the system   is to be followed  over the
Kelvin-Helmotz time scale,  $t_{K-H}  = GM^2/RL  \simeq 30M yr$  for a
solar-like star, the  Courant numbers we have  to deal with  is of the
order of  $10^{14}$.  These  rough estimates  \*  demonstrate that the
spatial resolution required   for  our  problem demands  an   implicit
numerical  scheme.  The process of  star formation is dominated by the
properties of radiative transfer in highly unsteady flows.  The reason
is that self-gravity alone can not lead to the  formation of a compact
object unless the  energy released by gravitational contraction beeing
radiated  away, from  the surface of  the   protostellar core.  Energy
transport in the core's interior,  towards the photosphere, is  mainly
due to   radiative  transfer in   an  optically  thick  plasma,    and
convection.  The photosphere  is a  narrow  layer where the  radiation
transport regime abruptly turns from the  diffusion regime to the free
streaming regime,  by which energy flowing  from the core by diffusion
and convection is  radiated  away.  Furthermore, during  the accretion
phase, the  photospere    is  surrounded  by  supercritical  radiating
accretion shock, where  the kinetic  energy of  the infalling  gas  is
essentially   converted into  outgoing radiation.   Since  most of the
total  luminosity  of  the  protostar  is produced within  this sharp,
optically  thin accretion shock, the calculation  of  the structure of
the radiative accretion  shock is a crucial  issue in  the modeling of
star   formation.  A correct  evaluation  of  the entropy extracted by
escaping radiation    from the accreting   material, deposited  on the
protostar  surface, is  also   the  key factor  which  determines  the
structure  of  the growing  stellar  core.  First, the entropy profile
determines the size of the quasistatic  core, and hence determines the
location of the protostar in the Hertzprung-Russell diagram by the end
of   the  accretion phase; second  the  entropy  gradient triggers the
highly efficient convective energy transport, in regions where $\nabla
s < 0$, where s is the specific entropy.
 
We  present in the following section  an example of the calculation of
the formation of a  primordial star, in  a slightly bound region of an
overall expanding universe.   The simple model presented here includes
only hydrodynamics,  gravitation and, of  course, radiative  transfer.
It is nevertheless a very challenging test for radiative transfer.  As
we will see, the opacity gradients are very large and there is a steep
transition from the diffusion to the free-streaming regime.

We start with a $10^6$ solar  mass gas cloud. The velocities are those
of  the homogeneous  expansion of  the universe  (i.e. $u(r)  = H\;r$,
where $H$ is the Hubble constant). The average density is the critical
density  and  the density  profile  is  smooth  with a  small  central
overdensity.  We start  our simulation at a redshift  of $100$ and the
amplitude of the initial perturbation  is adjusted so that the central
object forms  at a redshift  of $30$.  Initially the  radiative energy
density is that of the cosmic microwave background and the temperature
of the gas is equal to the temperature of the radiation.
 
\begin{figure}
\psfig{file=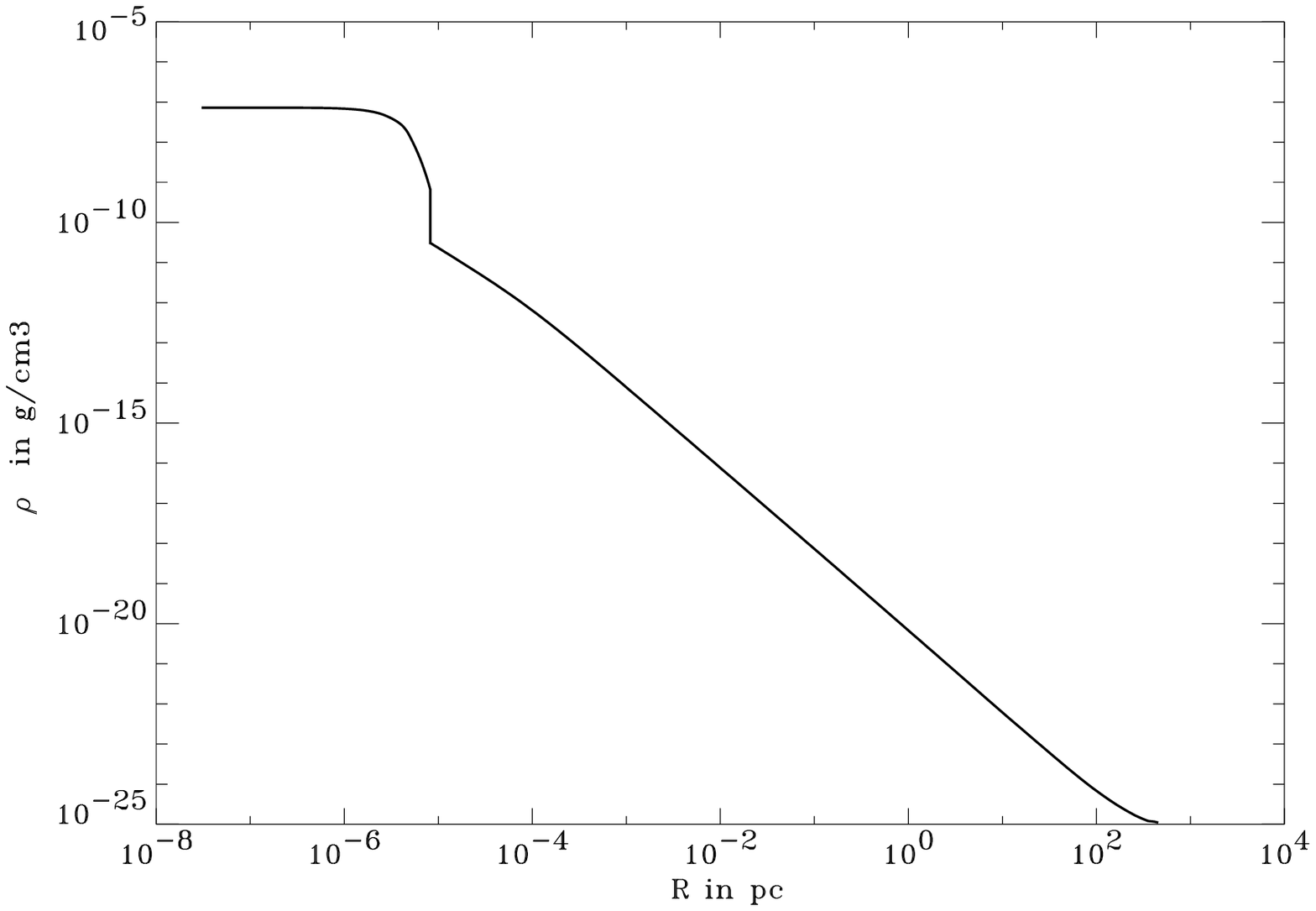,height=9.0cm}
\caption{ Density  profile of the  forming star. }
\label{prho}
\end{figure}

\begin{figure}
\psfig{file=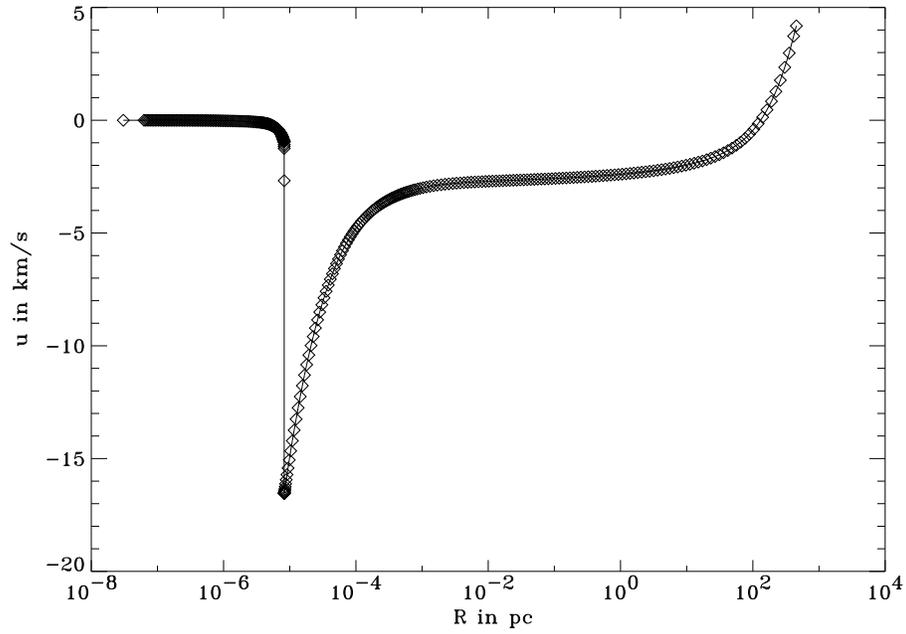,height=9.0cm}
\caption{ Velocity  profile of the  forming star.}
\label{pvit}
\end{figure}

\begin{figure}
\psfig{file=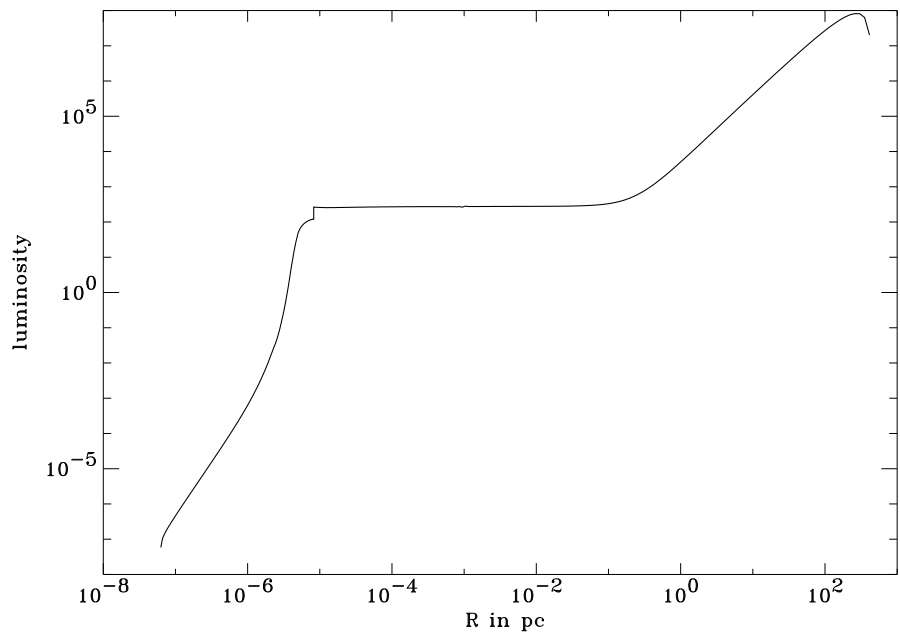,height=9.0cm}
\caption{  luminosity  profile of  the  forming  star.}
\label{plum}
\end{figure}

\begin{figure}
\psfig{file=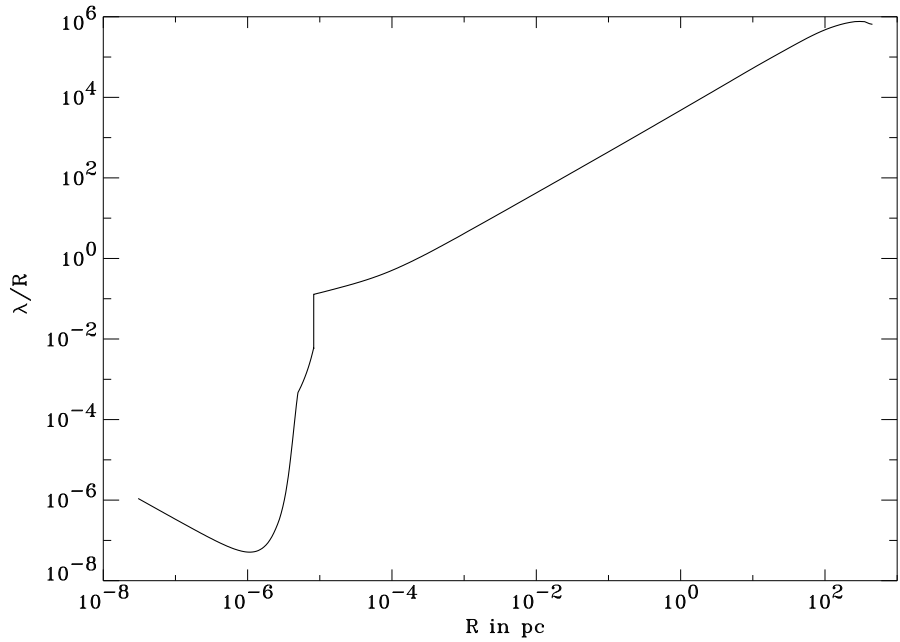,height=9.0cm}
\caption{ This plot represents the  ratio of the photon mean free path
over the radius as a function  of the radius. This ratio is very small
in the center where the  coupling between matter and radiation is very
strong, and  it is  very large in  the outer  region where the  gas is
transparent. This parameter varies over 10 orders of magnitude, with a
sharp  discontinuity at the  shock, where  the transition  between the
diffusion and the free streaming regime occurs. }
\label{plam}
\end{figure}

\begin{figure}
\psfig{file=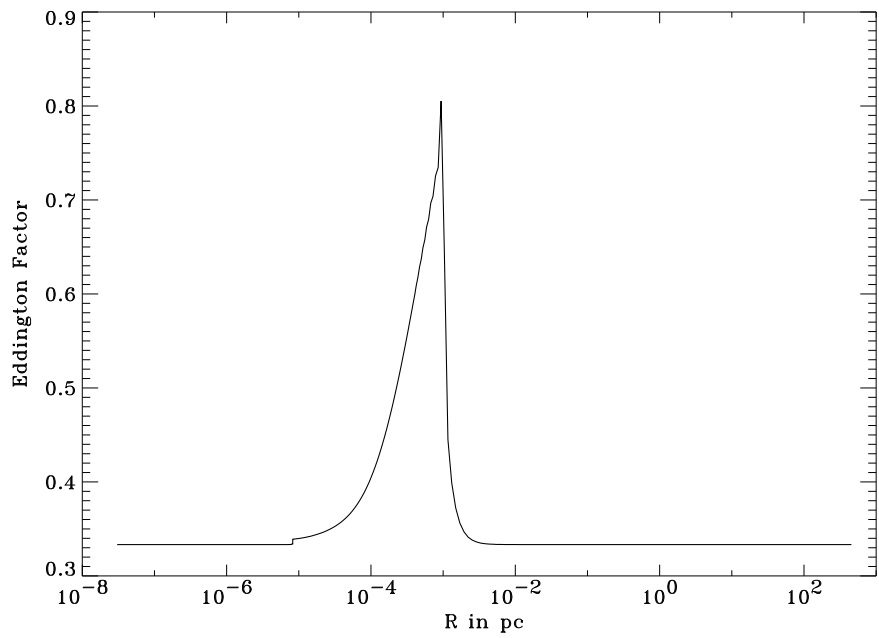,height=9.0cm}
\caption{Profil of the Eddington factor.}
\label{pfedd}
\end{figure}

\begin{figure}
\psfig{file=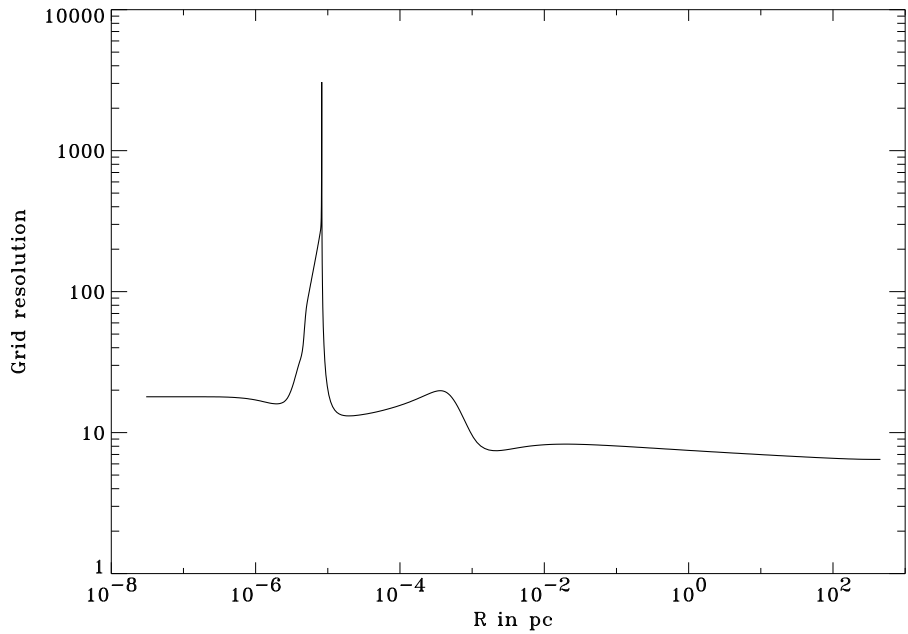,height=9.0cm}
\caption{ Profil of  the grid resolution. The grid  resolution is very
high at  the shock, where all  the gradients are very  large. With the
high resolution  it is  possible to see  the temperature spike  at the
shock (see  figure (\ref{ptemp})). The resolution is  also important in
the center to properly sample  the pressure gradient and near the peak
of the Eddington factor.}
\label{pn}
\end{figure}

\begin{figure}
\psfig{file=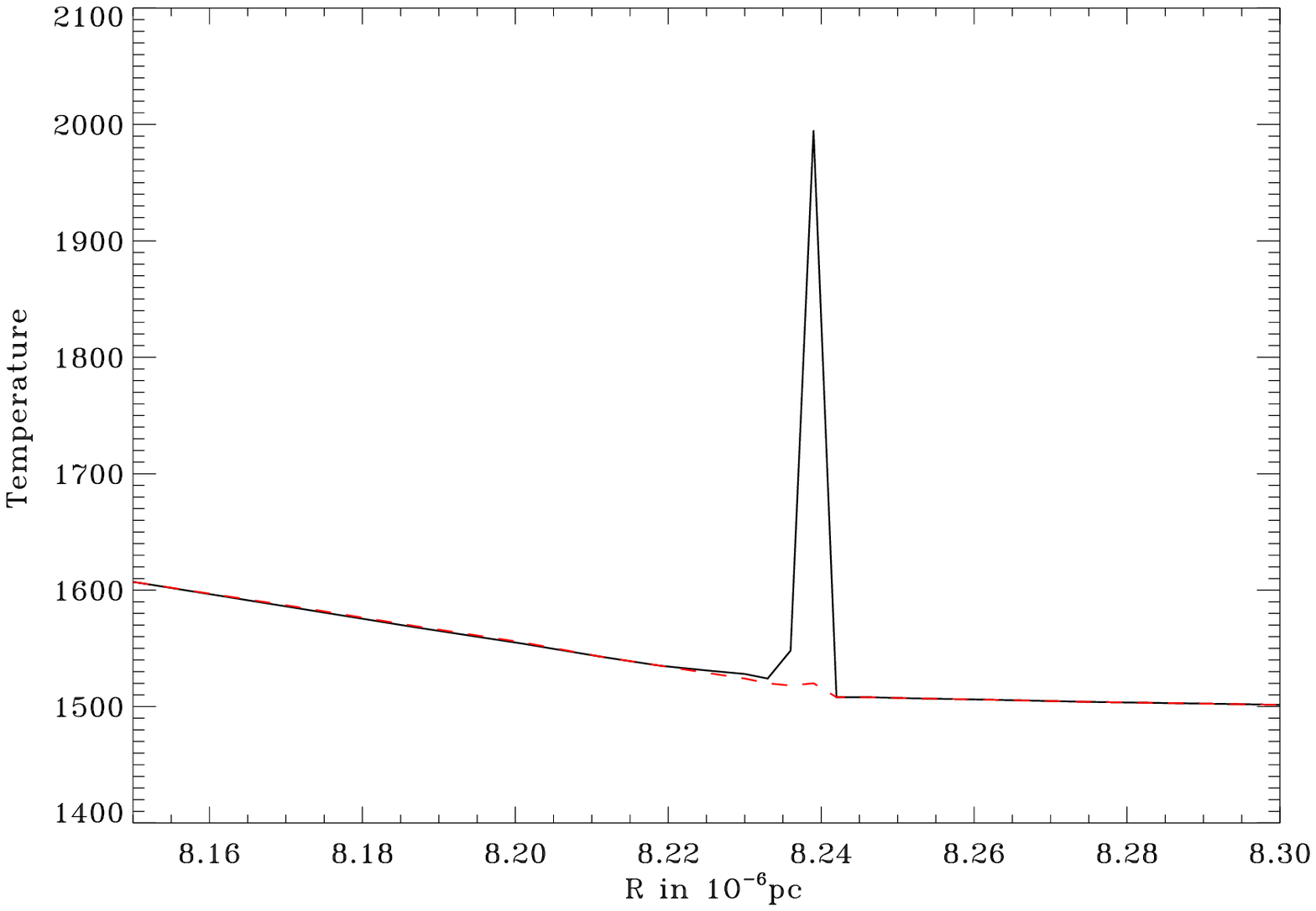,height=9.0cm}
\caption{Temperature profiles at  the shock. The gas is  heaten by the
shock and then very quickly radiates its internal energy. This results
in  the large  spike in  the  gas temperature  at the  shock which  is
associated to the jump in  luminosity.  On the contrary, the radiation
temperature  varies smoothy  through the  shock, it  only  changes its
slope.}
\label{ptemp}
\end{figure}

In a first phase the entire cloud is in expansion. Then, at a redshift
of $50$ the central part of the cloud starts  to collapse. The central
density increases, but the temperature remains almost constant because
the energy  is evacuated by radiation since  the photon mean free path
is much larger than the typical  scale of the forming structure. Next,
at a redshift  of about $30$,  the central  region is dense  enough to
become opaque to radiation. The  temperature and the pressure start to
grow and become  large enough to  prevent the collapse in the  central
region. Finally the infalling matter creates an accretion shock on the
central  core. The structure of the  flow when the central density has
reached $x.10^{-8} g/cm^3$   is presented in figures  (\ref{prho})  to
(\ref{ptemp}).

The density profile is plotted is figure  (\ref{prho}).  The center is
almost at the hydrostatic \* equilibrium and is surrounded by a slowly
contracting envelope.  The envelope  ends at the accretion shock where
the density jump is very important (about one  order of magnitude) due
to the fact that the gas radiates most of its energy, as is illusrated
by figures (\ref{plum}) and  (\ref{ptemp}).  One can notice  that  the
density contrast between the center and the outer accretion flow is of
more  than  18 decades  and that  the length  scale vary   of about 10
decades.

The  velocity  profile (figures  (\ref{pvit}))  shows  that the  outer
region of the cloud is still expanding, following the global expension
of the universe.  On the contrary, the inner region is gravitationally
bounded and is collapsing. The  accretion shock is clearly visible and
is sampled over only one  computational grid point.  Due to the moving
grid, the central part and the region of the shock have a high spatial
resolution.   The actual  resolution  profile can  be  seen on  figure
(\ref{pn}).  In the central part of the cloud, the radiation is in the
diffusion   limit   and  the   luminosity   slowly  increase   (figure
(\ref{plum})).   At  the shock  there  is  a  very sharp  increase  in
luminosity due  to the fact  that the gas  is heated at the  shock and
radiates most of  its energy.  The luminosity is  then constant except
for  the outer  region where  the luminosity  of the  cosmic microwave
background  starts   to  dominate.    As  is  illustrated   by  figure
(\ref{plam}), the radiation is in  the diffusion regime in the central
part of the cloud where the photon mean free path is much smaller than
the radius  while it  is clearly in  the free-streaming regime  in the
outer part.  The transition between these two regimes occurs in a very
sharp way at the shock.  Figure  (\ref{ptemp}) shows a zoom of the gas
and radiation temperatures profiles near  the shock.  The spike in the
gas temperature is a characteristic feature of supercritical radiative
shock and  a very  high spatial  resolution is needed  to resolve  it. 
Finally,  the profile  of the  Eddington factor  is plotted  on figure
(\ref{pfedd}).   The Eddington  factor is  equal to  one third  in the
central object,  which confirm the  fact that the diffusion  regime is
valid in this region. Then, there is a small jump at the shock and the
Eddington factor grows as one goes  away from the surface of the star. 
At large distance,  the light coming from the star  is blured into the
cosmic microwave background and the Eddington factor decreases back to
one third.   These large variations  of both the Eddington  factor and
the  opacity illustrate  the  interest  of the  $M_1$  model for  such
radiative shock.

\section{Conclusion}

We have presented  in this paper  a new numerical model for  radiative
transfer which gives satisfactory results in  a wide range of physical
conditions including both the diffusion and the free-streaming regimes
where the   anisotropy  of the  photons distribution  function  can be
large.  This method  is therefore well  suited  to modeled some  laser
experiments   and many astrophysical flows.  Works  are in progress to
incorporate  in the model a more  detailed description of the physical
processes  (multigroup  approach,   ionization,  ....)  and to further
develop this approach in a multidimensionnal framework.

\iftwocol}
\end{multicols}
\fi

\end{document}